\documentclass[12pt]{article}
\usepackage{epsfig}
\usepackage{amsmath}
\usepackage{color}
\usepackage{colortbl}
\usepackage{hhline}
\usepackage{amssymb}
\usepackage{times}
\usepackage{cite}
\usepackage{array}

\newlength{\dinwidth}
\newlength{\dinmargin}
\begin{document}  
%\linenumbers
% The rest
\newcommand{\pom}{{I\!\!P}}
\newcommand{\reg}{{I\!\!R}}
\newcommand{\slowpi}{\pi_{\mathit{slow}}}
\newcommand{\fiidiii}{F_2^{D(3)}}
\newcommand{\fiidiiiarg}{\fiidiii\,(\beta,\,Q^2,\,x)}
\newcommand{\n}{1.19\pm 0.06 (stat.) \pm0.07 (syst.)}
\newcommand{\nz}{1.30\pm 0.08 (stat.)^{+0.08}_{-0.14} (syst.)}
\newcommand{\fiidiiiful}{F_2^{D(4)}\,(\beta,\,Q^2,\,x,\,t)}
\newcommand{\fiipom}{\tilde F_2^D}
\newcommand{\ALPHA}{1.10\pm0.03 (stat.) \pm0.04 (syst.)}
\newcommand{\ALPHAZ}{1.15\pm0.04 (stat.)^{+0.04}_{-0.07} (syst.)}
\newcommand{\fiipomarg}{\fiipom\,(\beta,\,Q^2)}
\newcommand{\pomflux}{f_{\pom / p}}
\newcommand{\nxpom}{1.19\pm 0.06 (stat.) \pm0.07 (syst.)}
\newcommand {\gapprox}
   {\raisebox{-0.7ex}{$\stackrel {\textstyle>}{\sim}$}}
\newcommand {\lapprox}
   {\raisebox{-0.7ex}{$\stackrel {\textstyle<}{\sim}$}}
\def\gsim{\,\lower.25ex\hbox{$\scriptstyle\sim$}\kern-1.30ex%
\raise 0.55ex\hbox{$\scriptstyle >$}\,}
\def\lsim{\,\lower.25ex\hbox{$\scriptstyle\sim$}\kern-1.30ex%
\raise 0.55ex\hbox{$\scriptstyle <$}\,}
\newcommand{\pomfluxarg}{f_{\pom / p}\,(x_\pom)}
\newcommand{\dsf}{\mbox{$F_2^{D(3)}$}}
\newcommand{\dsfva}{\mbox{$F_2^{D(3)}(\beta,Q^2,x_{I\!\!P})$}}
\newcommand{\dsfvb}{\mbox{$F_2^{D(3)}(\beta,Q^2,x)$}}
\newcommand{\dsfpom}{$F_2^{I\!\!P}$}
\newcommand{\gap}{\stackrel{>}{\sim}}
\newcommand{\lap}{\stackrel{<}{\sim}}
\newcommand{\fem}{$F_2^{em}$}
\newcommand{\tsnmp}{$\tilde{\sigma}_{NC}(e^{\mp})$}
\newcommand{\tsnm}{$\tilde{\sigma}_{NC}(e^-)$}
\newcommand{\tsnp}{$\tilde{\sigma}_{NC}(e^+)$}
\newcommand{\st}{$\star$}
\newcommand{\sst}{$\star \star$}
\newcommand{\ssst}{$\star \star \star$}
\newcommand{\sssst}{$\star \star \star \star$}
\newcommand{\tw}{\theta_W}
\newcommand{\sw}{\sin{\theta_W}}
\newcommand{\cw}{\cos{\theta_W}}
\newcommand{\sww}{\sin^2{\theta_W}}
\newcommand{\cww}{\cos^2{\theta_W}}
\newcommand{\trm}{m_{\perp}}
\newcommand{\trp}{p_{\perp}}
\newcommand{\trmm}{m_{\perp}^2}
\newcommand{\trpp}{p_{\perp}^2}
\newcommand{\alp}{\alpha_s}
\newcommand{\alps}{\alpha_s}
\newcommand{\sqrts}{$\sqrt{s}$}
\newcommand{\LO}{$O(\alpha_s^0)$}
\newcommand{\Oa}{$O(\alpha_s)$}
\newcommand{\Oaa}{$O(\alpha_s^2)$}
\newcommand{\PT}{p_{\perp}}
\newcommand{\JPSI}{J/\psi}
\newcommand{\sh}{\hat{s}}
\newcommand{\uh}{\hat{u}}
\newcommand{\MP}{m_{J/\psi}}
\newcommand{\PO}{I\!\!P}
\newcommand{\xbj}{x}
\newcommand{\xpom}{x_{\PO}}
\newcommand{\ttbs}{\char'134}
\newcommand{\xpomlo}{3\times10^{-4}}  
\newcommand{\xpomup}{0.05}  
\newcommand{\dgr}{^\circ}
\newcommand{\pbarnt}{\,\mbox{{\rm pb$^{-1}$}}}
\newcommand{\gev}{\,\mbox{GeV}}
\newcommand{\WBoson}{\mbox{$W$}}
\newcommand{\fbarn}{\,\mbox{{\rm fb}}}
\newcommand{\fbarnt}{\,\mbox{{\rm fb$^{-1}$}}}
%
% Some useful tex commands
%
\newcommand{\qsq}{\ensuremath{Q^2} }
\newcommand{\gevsq}{\ensuremath{\mathrm{GeV}^2} }
\newcommand{\et}{\ensuremath{E_t^*} }
\newcommand{\rap}{\ensuremath{\eta^*} }
\newcommand{\gp}{\ensuremath{\gamma^*}p }
\newcommand{\dsiget}{\ensuremath{{\rm d}\sigma_{ep}/{\rm d}E_t^*} }
\newcommand{\dsigrap}{\ensuremath{{\rm d}\sigma_{ep}/{\rm d}\eta^*} }

\newcommand{\hdick}{\noalign{\hrule height1.4pt}}
\newcommand{\up}{\raisebox{1.7ex}[-1.7ex]}

\definecolor{indianred}{rgb}{0.5812,0.0665,0.0659} % IndianRed
\definecolor{orange}{rgb}{1.0,0.796875,0.2578125} % Orange

% Journal macro
\def\Journal#1#2#3#4{{#1} {\bf #2} (#3) #4}
\def\NCA{\em Nuovo Cimento}
\def\NIM{\em Nucl. Instrum. Methods}
\def\NIMA{{\em Nucl. Instrum. Methods} {\bf A}}
\def\NPB{{\em Nucl. Phys.}   {\bf B}}
\def\PLB{{\em Phys. Lett.}   {\bf B}}
\def\PRL{\em Phys. Rev. Lett.}
\def\PRD{{\em Phys. Rev.}    {\bf D}}
\def\ZPC{{\em Z. Phys.}      {\bf C}}
\def\EJC{{\em Eur. Phys. J.} {\bf C}}
\def\CPC{\em Comp. Phys. Commun.}

\hyphenation{DJANGOH}

\begin{titlepage}

\noindent

\begin{flushleft}
DESY 11-044    \hfill    ISSN 0418-9833 \\
March  2011                  \\
\end{flushleft}

\vspace*{3cm}

\begin{center}
\begin{Large}

{\bf Search for Lepton Flavour Violation at HERA\\}

\vspace{2cm}

H1 Collaboration

\end{Large}
\end{center}

\vspace{2cm}

\begin{abstract} \noindent

A search for second and third generation scalar and vector leptoquarks produced in
$ep$ collisions via the lepton flavour violating processes \mbox{$ep\rightarrow \mu  X$}
and \mbox{$ep\rightarrow \tau X$} is performed by the H1 experiment at HERA.
The full data sample taken at a centre-of-mass energy $\sqrt{s} = 319$~GeV is
used for the analysis, corresponding to an integrated luminosity of
$245$~pb$^{-1}$ of $e^{+}p$ and $166$~pb$^{-1}$ of $e^{-}p$ collision data.
No evidence for the production of such leptoquarks is observed in the H1 data.
Leptoquarks produced in $e^{\pm}p$ collisions with a coupling strength of
$\lambda=0.3$ and decaying with the same coupling strength to a muon-quark
pair or a tau-quark pair are excluded at $95\%$ confidence level up to
leptoquark masses of $712$~GeV and $479$~GeV, respectively.

\end{abstract}

\vspace{1.5cm}

\begin{center}
Submitted to \PLB
\end{center}

\end{titlepage}

\begin{flushleft}

F.D.~Aaron$^{5,48}$,           %BUCH-PD        11/06           Aaron               
C.~Alexa$^{5}$,                %BUCH-PD        06/06           Alexa               
V.~Andreev$^{25}$,             %LPI -PD        8/88            Andreev             
S.~Backovic$^{30}$,            %PODG-PD        03/02           Backovic            
A.~Baghdasaryan$^{38}$,        %YERE-PD        09/03           Baghdasaryana       
S.~Baghdasaryan$^{38}$,        %YERE-ST        02/10           Baghdasaryans       
E.~Barrelet$^{29}$,            %PARI-PD        11/99           Barrelet            
W.~Bartel$^{11}$,              %DESY-PD        8/88            Bartel              
K.~Begzsuren$^{35}$,           %ULBA-PD        04/06           Begzsuren           
A.~Belousov$^{25}$
% ...

%-- H1 Institutes 
\bigskip{\it
 $ ^{1}$ I. Physikalisches Institut der RWTH, Aachen, Germany \\
% ...
$ ^{5}$ National Institute for Physics and Nuclear Engineering (NIPNE) ,
          Bucharest, Romania$^{ m}$ \\
% ...
$ ^{11}$ DESY, Hamburg, Germany \\
% ...
$ ^{25}$ Lebedev Physical Institute, Moscow, Russia$^{ e}$ \\
% ...
$ ^{29}$ LPNHE, Universit\'e Pierre et Marie Curie Paris 6,
          Universit\'e Denis Diderot Paris 7, CNRS/IN2P3, Paris, France \\
 $ ^{30}$ Faculty of Science, University of Montenegro,
          Podgorica, Montenegro$^{ n}$ \\
% ...
$ ^{35}$ Institute of Physics and Technology of the Mongolian
          Academy of Sciences, Ulaanbaatar, Mongolia \\
% ...
$ ^{38}$ Yerevan Physics Institute, Yerevan, Armenia \\
% ...
$ ^{48}$ Also at Faculty of Physics, University of Bucharest,
          Bucharest, Romania \\
% ...
$ ^e$ Supported by the Deutsche Forschungsgemeinschaft \\
% ...
$ ^m$ Supported by the Romanian National Authority for Scientific Research
      under the contract PN 09370101 \\
 $ ^n$ Partially Supported by Ministry of Science of Montenegro,
      no. 05-1/3-3352 \\
}

\end{flushleft}

\newpage
%%%%%%%%%%%%%%%%%%%%%%%%%%%%%%%%%%%%%%%%%%%%%%%%%%%%%%%%%%%%
\section{Introduction}
\label{sec:intro}

The $ep$ collisions at HERA provide a unique testing ground to search for new
particles coupling directly to a lepton and a quark.
An example of such particles are leptoquarks (LQs), colour triplet bosons which
are a generic prediction of grand unified theories~\cite{Pati:1974yy}, composite
models~\cite{Schrempp:1984nj}, technicolour~\cite{Dimopoulos:1979es}
and supersymmetry with $R$-parity violation~\cite{Nilles:1983ge}.
In the Standard Model (SM) particle interactions conserve lepton flavour, and if this
property is extended to LQ models, any such particles produced at HERA would decay
exclusively into a quark and a first generation lepton, namely an
electron\footnote{In this letter the term ``electron'' is used generically to refer to both
electrons and positrons, if not otherwise stated.} or a neutrino.
Dedicated searches have been performed at HERA for such leptoquarks, where the
SM expectation is dominated by neutral current (NC) and charged current (CC) deep
inelastic scattering (DIS) background\cite{Adloff:1999tp,Aktas:2005pr,Chekanov:2003af}.

%%%

The introduction of lepton flavour violation (LFV) to leptoquark models would mean
the processes $ep \rightarrow \mu X$ or $ep \rightarrow \tau X$, mediated by the
exchange of a leptoquark, would be observable at HERA with final states containing a
muon or the decay products of a tau lepton in combination with a hadronic system $X$.
Searches for such signatures have been performed at HERA and limits on LFV leptoquark
production have been derived~\cite{Adloff:1999tp,Aktas:2007ji,Chekanov:2005au}.
In this paper a search for LFV phenomena is performed using $e^{\pm}p$ collision data at a
centre-of-mass energy $\sqrt{s} = 319$~GeV, recorded during the years 1998-2007
by the H1 experiment at HERA.
The corresponding integrated luminosity of $245$~pb$^{-1}$ for $e^{+}p$ collisions and
$166$~pb$^{-1}$ for $e^{-}p$ collisions represents an increase in size of the data sample
with respect to the previous publication by a factor of $3$ and $12$, respectively.
Data collected from 2003 onwards were taken with a longitudinally polarised lepton beam,
with polarisation typically at a level of $35\%$.
The presented results supersede those derived in previous searches for lepton flavour
violating leptoquarks by the H1 experiment\cite{Aktas:2007ji}.

%%%%%%%%%%%%%%%%%%%%%%%%%%%%%%%%%%%%%%%%%%%%%%%%%%%%%%%%%%%%
\section{Leptoquark Phenomenology}
\label{sec:theory}

The phenomenology of LQs at HERA is discussed in detail elsewhere~\cite{Adloff:1999tp}.
In the framework of the Buchm\"uller-R\"uckl-Wyler (BRW) effective model \cite{BRW}, LQs are
classified into $14$ types~\cite{14LQ} with respect to the quantum numbers spin $J$, weak isospin $I$ and
chirality $C (= L,R)$.
Scalar ($J=0$) LQs are denoted as $S_{I}^{C}$ and vector ($J=1$) LQs are
denoted $V_{I}^{C}$ in the following.
LQs with identical quantum numbers but different weak hypercharge are distinguished
using a tilde, for example $V_0^R$ and $\tilde{V}_0^R$.
Some LQs, namely $S_0^L$, $S_{1}^L$, $V_0^L$ and $V_{1}^L$, may decay to a neutrino-quark pair
resulting in the branching fraction for decays into charged leptons
\mbox{$\beta_\ell\!=\!\Gamma_{\ell q}/(\Gamma_{\ell q}+\Gamma_{\nu_\ell q})\!=0.5$}.
Since neutrino flavours cannot be distinguished with the H1 experiment such final states
are not included in this analysis.

%%%

Leptoquarks carry both lepton ($L$) and baryon ($B$) quantum numbers, and the fermion number
\mbox{$F\!=\!L\!+\!3\,B$} is assumed to be conserved.
Leptoquark processes proceed directly via $s$-channel resonant LQ production or indirectly
via $u$-channel virtual LQ exchange.
For LQ masses $m_{\rm LQ}$ well below $\sqrt{s}$, the $s$-channel production of
$F = 2$ ($F = 0$) LQs in $e^-p$ ($e^+p$) collisions dominates.
However, for LQ masses above $319\,{\rm GeV}$, both the $s$ and $u$-channel processes
are important such that both $e^-p$ and $e^+p$ collisions have similar sensitivity
to LQs with $F = 2$ and LQs with $F = 0$.

%%%

The BRW model assumes lepton flavour conservation, although a general extension of this model
allows for the decay of LQs to final states containing a quark and a lepton of a different flavour,
that is a muon or tau lepton.
Non-zero couplings $\lambda_{eq_i}$ to an electron-quark pair and $\lambda_{\mu q_j}$
($\lambda_{\tau q_j}$) to a muon(tau)-quark pair are assumed.
The indices $i$ and $j$ represent quark generation indices, such that $\lambda_{eq_i}$ denotes the
coupling of  an electron to a quark of generation $i$, and $\lambda_{\ell q_j}$ is the coupling 
of the outgoing lepton (where $\ell = \mu$ or $\tau$) to a quark of generation $j$.
An overview of this extended model for the LQ coupling to $u$ and $d$ quarks is provided
elsewhere~\cite{Aktas:2007ji}.

%%%

Events with LQs are generated using the LEGO \cite{lego} event generator with the CTEQ5L
parametrisation~\cite{Pumplin:2002vw} of the parton distribution functions of the proton. 
The LQ signal expectation is calculated as a function of the LQ type, mass, coupling constant and the 
branching ratio $\beta$ to a given charged lepton flavour, where:
\[
\beta = \beta_{\ell} \times \beta_{LFV} %for $LQ\rightarrow\mu(\tau)q$.
~~\text{with}
~~\beta_{LFV}=\frac{\Gamma_{\mu(\tau)q}}{\Gamma_{\mu(\tau)q}+\Gamma_{eq}}~~\text{and}
~~\Gamma_{\ell q}=m_{{\rm LQ}}\lambda^2_{\ell q}\times 
  \begin{cases}
    \frac{1}{16\pi} & \text{scalar LQ}\\
    \frac{1}{24\pi} & \text{vector LQ}
  \end{cases}
\]
where \mbox{$\Gamma_{\ell q}$} denotes the partial LQ decay width for the decay
to a lepton \mbox{$\ell=e,\mu,\tau$} and a quark $q$.
In order to avoid the need to generate many Monte Carlo (MC) samples at each leptoquark
mass, coupling and branching ratio, a weighting technique is used to provide predictions across
the full range of LQ production parameters~\cite{Aktas:2007ji}.
%

%%%

Leptoquarks with couplings to the first and the second lepton generation
may decay to a muon and a quark, leading to an event topology with an isolated
high transverse momentum $P_T$ muon back-to-back to a hadronic system in
the transverse plane.
Leptoquarks with couplings to the first and the third lepton generation may decay
to a tau and a quark.
Tau leptons are identified in this analysis using the muonic and one-prong hadronic
decays of the tau.
In both cases, the tau decay results in missing transverse momentum in the event due to
the escaping neutrinos. 
Previous LFV leptoquark analyses also examined $\tau \rightarrow e X$ decays 
and three-prong hadronic tau decays~\cite{Aktas:2007ji}, for which the background
from SM processes is large~\cite{lindfeld}.
Given the increase in data luminosity with respect to the previous publication,
a correspondingly large increase in the SM background is observed, which limits 
the sensitivity of these decay channels and they are therefore not included in the
presented analysis.

%%%%%%%%%%%%%%%%%%%%%%%%%%%%%%%%%%%%%%%%%%%%%%%%%%%%%%%%%%%%
\section{Standard Model Background Processes}
\label{sec:sm}

Several SM processes may mimic the LQ signal.
The main SM background contribution is from photoproduction events, in which a
hadron is wrongly identified as a muon or a narrow hadronic jet fakes the signature
of the hadronic tau decay.
Similarly, the scattered electron in NC DIS events may also be misinterpreted as the
one-prong hadronic tau decay jet.
Smaller SM background contributions arise from events exhibiting intrinsic
missing transverse momentum (for example CC DIS), events containing high $P_{T}$
leptons (such as lepton pair production, particularly inelastic muon-pair events if one
muon is unidentified) or events with both of these features (real $W$ production with
leptonic decay).

%%%

The RAPGAP~\cite{Jung:1993gf} event generator, which implements the Born level,
QCD Compton and boson-gluon fusion matrix elements, is used to model inclusive NC DIS
events.
The QED radiative effects arising from real photon emission from both the incoming and outgoing
electrons are simulated using the HERACLES~\cite{Kwiatkowski:1990es} program.
Direct and resolved photoproduction of jets and prompt photon production are simulated using
the PYTHIA~\cite{Sjostrand:2000wi} event generator, which is based on Born level scattering matrix
elements.
In RAPGAP and PYTHIA, jet production from higher order QCD radiation is simulated using leading
logarithmic parton showers and hadronisation is modelled with Lund string fragmentation~\cite{lund}. 
Inclusive CC DIS events are simulated using the DJANGOH~\cite{Schuler:yg} program, which includes first
order leptonic QED radiative corrections based on HERACLES.
The production of two or more jets in DJANGOH is accounted for using the colour dipole
model~\cite{Lonnblad:1992tz}. 
The leading order MC prediction of processes with two or more high transverse momentum jets in
NC DIS, CC DIS and photoproduction is scaled by a factor of $1.2$ to account for the incomplete
description of higher orders in the MC generators~\cite{Adloff:2002au,Aktas:2004pz}. 
Contributions arising from the production of single $W$ bosons and multi-lepton events are modelled
using the EPVEC~\cite{Baur:1991pp}  and GRAPE~\cite{Abe:2000cv} event generators, respectively. 
The uncertainties on the SM background predictions are described in section~\ref{sec:sys}.

%%%

Generated events are passed through a GEANT~\cite{Brun:1987ma} based simulation
of the H1 apparatus,  which takes into account the running conditions of the data taking.
Simulated events are reconstructed and analysed using the same program chain as is used
for the data.

%%%%%%%%%%%%%%%%%%%%%%%%%%%%%%%%%%%%%%%%%%%%%%%%%%%%%%%%%%%%
\section{Experimental Conditions}
\label{sec:exp}

A detailed description of the H1 experiment can be found elsewhere~\cite{Abt:h1}.
Only the detector components relevant to this analysis are briefly described here.
A right-handed Cartesian coordinate system is used with the origin at the nominal primary
$ep$ interaction vertex. 
The proton beam direction defines the positive $z$ axis (forward direction).
The polar angle $\theta$ and the transverse momenta $P_T$ of all particles are defined with
respect to this axis.
The azimuthal angle $\phi$ defines the particle direction in the transverse plane.
The pseudorapidity is defined as $\eta=-\ln {\tan {\frac{\theta}{2}}}$.

%%%

The Liquid Argon (LAr) calorimeter~\cite{Andrieu:1993kh} covers the polar angle range
\mbox{$4^\circ < \theta < 154^\circ$} with full azimuthal acceptance.
The energies of electromagnetic showers are measured in the LAr with a precision of
\mbox{$\sigma (E)/E \simeq 11\%/ \sqrt{E/\mbox{GeV}} \oplus 1\%$} and hadronic
energy depositions with \mbox{$\sigma (E)/E \simeq 50\%/\sqrt{E/\mbox{GeV}} \oplus 2\%$},
as determined in test beam measurements~\cite{Andrieu:1993tz,Andrieu:1994yn}.
A lead-scintillating fibre calorimeter (SpaCal)~\cite{SpaCal} covering the backward region
\mbox{$153^\circ < \theta < 178^\circ$} completes the measurement of charged and neutral
particles.
For electrons a relative energy resolution of
\mbox{$\sigma (E)/E \simeq 7\%/\sqrt{E/\mbox{GeV}} \oplus 1\%$} is reached, as determined
in test beam measurements~\cite{SpaTestBeam}. 
The central \mbox{($20^\circ < \theta < 160^\circ$)}  and forward
\mbox{($7^\circ < \theta < 25^\circ$)}  inner tracking detectors are used to measure charged
particle trajectories and to reconstruct the interaction vertex.
The measured trajectories fitted to the interaction vertex are referred to as tracks
in the following.
The LAr calorimeter and inner tracking detectors are enclosed in a superconducting
magnetic coil with a field strength of $1.16$~T.
From the curvature of charged particle trajectories in the magnetic field, the central
tracking system provides transverse momentum measurements with a resolution of 
\mbox{$\sigma_{P_T}/P_T = 0.005 P_T / \mbox{GeV} \oplus 0.015$}~\cite{Kleinwort:2006zz}.
The return yoke of the magnetic coil is the outermost part of the detector and is equipped with
streamer tubes forming the central muon detector \mbox{($4^\circ < \theta < 171^\circ$)}. 
In the very forward region of the detector \mbox{($3^\circ < \theta < 17^\circ$)} a set of drift
chambers detects muons and measures their momenta using an iron toroidal magnet.
The luminosity is determined from the rate of the Bethe-Heitler process
$ep \rightarrow ep \gamma$, measured using a photon detector located close to the beam
pipe at $z=-103~{\rm m}$, in the backward direction.

%%%

Lepton flavour violating processes usually exhibit an imbalance in the measured calorimetric
transverse momentum,  $P_{T}^{\rm calo}$, due to either the presence of a minimally ionising
muon in $\mu X$ final states or the escaping neutrino(s) from tau decays in $\tau X$ events. 
The LAr calorimeter provides the main trigger in this analysis.
The trigger efficiency is about $60\%$ for events with a transverse momentum imbalance
measured in the calorimeter of $12$~GeV, rising to about $90\%$ for an imbalance
of $25$~GeV~\cite{Adloff:2003uh}.
Events are also triggered by hadronic jets in the LAr calorimeter, with a trigger efficiency
above $95\%$ for a jet transverse momentum $P_{T}^{\rm jet} > 20$~GeV and
almost $100\%$ for $P_{T}^{\rm jet} > 25$~GeV\cite{peez}.
For di-jet events with a scalar sum of the transverse energy in the event
$E_{T} > 30$~GeV, the trigger efficiency is greater than $98\%$~\cite{Aktas:2006qe}.

%%%

In order to remove events induced by cosmic rays and other non-$ep$ background,
the event  vertex is required to be reconstructed within $\pm 35$~cm in $z$ of the average
nominal interaction point.
In addition, topological filters and timing vetoes are applied.

%%%%%%%%%%%%%%%%%%%%%%%%%%%%%%%%%%%%%%%%%%%%%%%%%%%%%%%%%%%%
\section{Particle Identification and Event Selection}
\label{sec:partid}

Electromagnetic particle (electron and photon) candidates are identified as compact and isolated 
clusters of energy in the electromagnetic part of the LAr calorimeter.
Electron candidates are defined as electromagnetic particle candidates with an associated track.
Identification of muon  candidates is based on a track in the inner tracking detectors,
associated to a signal in the muon system.
Tracks and calorimeter deposits not identified as originating from isolated electromagnetic particles
or muons are combined into cluster-track objects to reconstruct the hadronic final state~\cite{h1lt}.
Jets are reconstructed from these objects using an inclusive $k_{T}$
algorithm~\cite{Ellis:1993tq,Catani:1993hr} with a minimum $P_{T}$ of $4$ GeV and a distance
parameter $R=1.0$.
The missing transverse momentum $P_{T}^{\rm miss}$, which may indicate the presence of neutrinos
in the final state, is derived from all reconstructed particles in the event.
The LQ kinematics are reconstructed using the double angle method~\cite{DAmeth}.
The direction of the detected lepton and the hadronic final state are used to reconstruct
the Bjorken scaling variable $x$ and subsequently the LQ mass $m_{\rm LQ} =\sqrt{xs}$.

%%%%%%%%%%%%%%%%%%%%%%%%%%%%%%%%%%%%%%%%%%%%%%%%%%%%%%%%%%%%
\subsection{Search for second generation leptoquarks}
\label{sec:secondgen}

An initial sample of events with muons and jets is selected by requiring at least one
$P_{T}^{\mu} > 8$~GeV muon in the polar angular range $10^{\circ} < \theta_{\mu} < 120^{\circ}$
and at least one jet.
In addition, $P_{T}^{\rm calo}$ is required to be greater than $12$~GeV.
After this selection, $996$ events are observed in the data, in good agreement with the
SM prediction of $978 \pm 187$, where the uncertainty includes the statistical and
systematic errors (see section~\ref{sec:sys}).

%%%

Events with at least one isolated muon are then selected, which is done by
requiring the angular distance, $D=\sqrt{(\Delta\eta)^2+(\Delta\phi)^2}$, of the muon
to the nearest track and to the nearest jet to be greater
than $0.5$ and $1.0$, respectively.
In addition, an isolated muon may have no more than $5$~GeV deposited in the LAr
calorimeter within a cylinder centred on the muon track direction of radius $35$~cm ($75$~cm)
in the electromagnetic (hadronic) section.
The muon isolation requirements reduce the number of selected data events to
$220$, compared to a SM prediction of $218 \pm 48$.

%%%

The NC DIS background is further suppressed by increasing the cut on the calorimetric
momentum imbalance to $P_{T}^{\rm calo}>25$~GeV, which implicitly increases the minimum
muon transverse momentum, and by rejecting events with identified isolated electrons.
To reduce the muon-pair SM background, exactly one isolated muon is required,
as expected in LFV LQ signal events.
The back-to-back event topology in the azimuthal plane is also exploited to
remove the SM background and the difference between the azimuthal angle of the
hadronic system and the muon $\Delta\phi_{\mu-X}$ is required to be greater
than $170^{\circ}$.
As the majority of the energy deposited in the calorimeter is due to the hadronic final
state, signal events tend to exhibit an azimuthal imbalance when considering the
calorimeter measurement alone.
Therefore, a requirement of $V_{\rm ap}/V_{\rm p}<0.3$ is also employed, where
$V_{\rm ap}/V_{\rm p}$ is defined as the ratio of the anti-parallel to parallel projections of
all energy deposits in the calorimeter with respect to the direction of
$P_{T}^{\rm calo}$~\cite{Adloff:1999ah}.
After these selection cuts, the data sample is reduced to $6$ events, compared to
a SM prediction of $7.5 \pm 1.8$.

%%%

To exploit the longitudinal balance of the event, a requirement on the sum of
the energy and longitudinal momentum of all detected particles $i$ in the event
$\Sigma_{i}(E^{i} - P_{z}^{i}) > 40$~GeV is applied.
In the case of signal events this quantity is expected to be around
$2E_{e}^{0} = 55.2$~GeV, where $E_{e}^{0}$ is the electron beam energy.
However, for the remaining SM background after the above event selection the
scattered electron or some other backward going final state particle is
typically undetected, resulting in significantly lower values of
$\Sigma_{i}(E^{i} - P_{z}^{i})$.
In order to improve the resolution, which is poor for very high $P_{T}$ muons due to
the small curvature of the track, the transverse momentum of the muon is calculated
from the hadronic system, \mbox{$\vec{P}_{T}^{\mu} = -\vec{P}_{T}^{X}$} and
the muon track direction is used to reconstruct the longitudinal component
$P_{z}^{\mu}$ and energy $E^{\mu}$ used in the $(E - P_{z})$ sum~\cite{panagoulias}.

%%%

The cut on $\Sigma_{i}(E^{i} - P_{z}^{i}) $ removes five of the remaining data events, so that
one event is observed in the final selection of the analysis of $\mu X$ final states,
which compares well to the SM prediction of $2.0 \pm 0.4$, where the largest contribution
comes from muon-pair events.
The presented analysis has a lower background contamination and an improved selection
efficiency with respect to the previous H1 publication.
The selection efficiency typically ranges from $75\%$ for LQs masses of
around $150$~GeV to $65\%$ for LQ masses above $300$~GeV, representing
an improvement of an additional $15-25\%$ with respect to the previous
publication~\cite{Aktas:2007ji}.

%%%%%%%%%%%%%%%%%%%%%%%%%%%%%%%%%%%%%%%%%%%%%%%%%%%%%%%%%%%%
\subsection{Search for third generation leptoquarks}
\label{sec:thirdgen}

In the search for third generation leptoquarks, tau leptons are identified in this analysis
using the muonic and one-prong hadronic decays of the tau.

%%%

Muonic tau decays $\tau\rightarrow \mu\nu_\mu\nu_\tau$ result in final states similar to the
high $P_{T}$ muon signatures described in section~\ref{sec:secondgen}.
The same selection cuts are therefore applied in this channel.
To account for possible effects due to different muon kinematics resulting from the tau decay,
the selection efficiency was studied in an LFV MC signal sample with
a $\tau X$ final state and a subsequent muonic tau decay. 
The selection efficiency in this channel is up to $70\%$ at leptoquark masses of
around $150$~GeV and about $55\%$ for LQ masses above $300$~GeV, which
represents a similar  level of improvement with respect to the previous
publication~\cite{Aktas:2007ji} as observed in the second generation search described
in section~\ref{sec:secondgen}.

%%%

The one-prong hadronic decay of the tau leads to a high $P_T$, narrow ``pencil-like''
jet, so that the typical LFV signal event topology is a di-jet event.
An initial event sample for the analysis of this decay channel is formed by selecting
events with at least two jets in the polar angle range
$5^{\circ} < \theta^{\rm jet} < 175^{\circ}$ and with
$P_{T}^{\rm jet 1} > 20$~GeV and $P_{T}^{\rm jet 2} > 15$~GeV.
This results in a large di-jet sample of approximately $2.2 \cdot 10^{5}$
events, which is consistent with the SM prediciton of $(2.6 \pm 0.5) \cdot 10^{5}$,
where the main contribution comes from photoproduction.

%%%

A tau jet is characterised by a narrow energy deposit in the calorimeter and
a low track multiplicity within the identification cone of the jet.
Tau jet candidates are identified in the di-jet sample, where the candidate
is required to be in the polar angle range $20^{\circ} < \theta^{\rm jet} < 120^{\circ}$
and has a maximum jet radius $R_{\rm jet}$ of $0.12$~\cite{Aktas:2006fc}.
The jet radius is used as a measure of the collimation of the jet and is calculated as: 
\mbox{$R_{\rm jet}=\frac{1}{E_{\rm jet}}\sum_{h}E_{h}\sqrt{\Delta\eta({\rm jet},h)^{2}+\Delta\phi({\rm jet},h)^{2}}$},
where $E_{\rm jet}$ is the total jet energy and the sum runs over all jet daughter
hadronic final state particles of energy $E_{h}$.
At least one track with $P_{T}$ larger than $2$~GeV not associated with an identified
electron or muon is required within the jet radius of the tau jet candidate.
Approximately $3 \cdot 10^{4}$ tau jet candidates are identified in the di-jet sample.

%%%

The undetected neutrinos from tau lepton decays result in an overall $P_{T}$ imbalance
and therefore a minimum missing transverse momentum $P_{T}^{\rm miss} > 12$~GeV is
required.
Events with only one tau jet are then selected, which is required to be isolated from
tracks and other jets in the event by a distance $D > 1.0$.
A track multiplicity of one is required in a cone of radius $R = 1.0$ around the
jet axis.
Tau jets with additional track segments not fitted to the event vertex within a
cone of radius $R = 0.3$ around the jet axis are also rejected~\cite{Aaron:2009wp}.
To reject purely electromagnetic jets, a maximum of $90\%$ of the jet energy
may be recorded in the electromagnetic part of the calorimeter.
The resulting selection contains $104$ data events compared to a
SM prediction of $116 \pm 16$.

%%%

Further cuts are then applied to reduce the remaining SM background.
The hadronic transverse momentum $P_{T}^{X}$ is required to be
larger than $30$~GeV and the acoplanarity between the tau jet and $X$ system in the
transverse plane $\Delta\phi_{\tau - X}$ is required to be greater than $160^{\circ}$.
Note that for the analysis of the hadronic tau decay channel, the tau jet is subtracted from the
inclusive hadronic final state to obtain the four-vector of the remaining hadronic system $X$.
Analogous to the muon channel, a cut of $\Sigma_{i}(E^{i} - P_{z}^{i}) > 40$~GeV is applied
to exploit the longitudinal balance of the event.
Similarly to the muon channel, only the direction of the tau jet is used in the sum,
and the transverse momentum of the $X$ system is employed to determine the
tau jet four-vector. 
Electrons entering inactive regions of the electromagnetic section of the LAr calorimeter
may fake the tau jet signature and therefore these regions are excluded from the
analysis~\cite{panagoulias}.

%%%

In the analysis of $\tau X$ final states where the tau lepton decays hadronically,
$6$ events are observed in the data, in good agreement with the SM prediction of 
$8.2 \pm 1.1$, where the main SM contribution is from remaining NC DIS events.
The selection efficiency in the one-prong hadronic tau decay channel ranges
from $18\%$ for leptoquark masses in the range $150$-$200$ GeV to $12\%$ for
masses above $300$~GeV.

%%%%%%%%%%%%%%%%%%%%%%%%%%%%%%%%%%%%%%%%%%%%%%%%%%%%%%%%%%%%
\section{Systematic Uncertainties}
\label{sec:sys}

The following experimental systematic uncertainties are considered in the search for
second generation leptoquarks:
the scale uncertainty on the transverse momentum of high $P_{T}$ muons is $2.5\%$
and the uncertainty on the muon polar angle measurement $3$~mrad~\cite{Aaron:2008jh};
the muon identification efficiency has an error of $5\%$ in the region
$\theta^{\mu} > 12.5^{\circ}$ and $15\%$ in the forward region~\cite{Aaron:2009wp};
the hadronic energy scale is known within $2\%$ and the uncertainty on the hadronic
polar angle measurement is $10$~mrad~\cite{peez}.
In the search for third generation leptoquarks an uncertainty on the
description of the jet radius $R_{\rm jet}$ is included in the analysis by varying the
cut value of $0.12$ by $10\%$.
All other experimental systematic uncertainties in the tau channel are included
in the model uncertainties described below~\cite{Aaron:2009wp}.
In both searches, the uncertainty on the trigger efficiency is $2$-$3\%$
and the uncertainty on the luminosity measurement is $3\%$.

%%%

The effects of these systematic uncertainties on the signal and the expected SM background 
are evaluated by shifting the relevant quantities in the MC simulation by their uncertainty 
and adding all resulting variations in quadrature.

%%%

Additional model uncertainties are attributed to the normalisation uncertainties
in the analysis phase space of the SM MC generators described in section~\ref{sec:sm}.
These model uncertainties are estimated from control analyses in an extended
phase space relevant to the search signature~\cite{Aaron:2009wp}.
In the analysis of $\mu X$ final states, the contributions from RAPGAP (NC DIS), PYTHIA
(photoproduction) and GRAPE (lepton-pair production) are each attributed a systematic
error of $30\%$, which is increased to $50\%$ for the period 1998-2000~\cite{panagoulias}.
The contribution from DJANGOH (CC DIS) in events with isolated muons is attributed
an uncertainty of $50\%$.
In the analysis of $\tau X$ final states where the tau lepton decays hadronically,
the contribution from RAPGAP, PYTHIA, DJANGOH and GRAPE are attributed systematic
uncertainties of $15\%$, $20\%$, $20\%$ and  $30\%$, respectively.
The theoretical uncertainty of $15\%$ is used for all predicted contributions
from EPVEC ($W$ production)~\cite{Baur:1991pp}.

%%%

The total error on the SM prediction is determined by adding the MC statistical
error to the effects of all model and experimental systematic uncertainties
in quadrature.

%%%

The main theoretical uncertainty on the signal cross section originates from the
parton densities.
This uncertainty is estimated to be $5\%$ for LQs coupling to up-type quarks
and varies between $7\%$ at low masses and $30\%$ at masses around
$290$~GeV for LQs coupling to down-type quarks~\cite{Aktas:2005pr}.

%%%%%%%%%%%%%%%%%%%%%%%%%%%%%%%%%%%%%%%%%%%%%%%%%%%%%%%%%%%%
\section{Results}
\label{sec:res}

\begin{figure}[ht] 
  \begin{center}
   \includegraphics[width=0.7\textwidth]{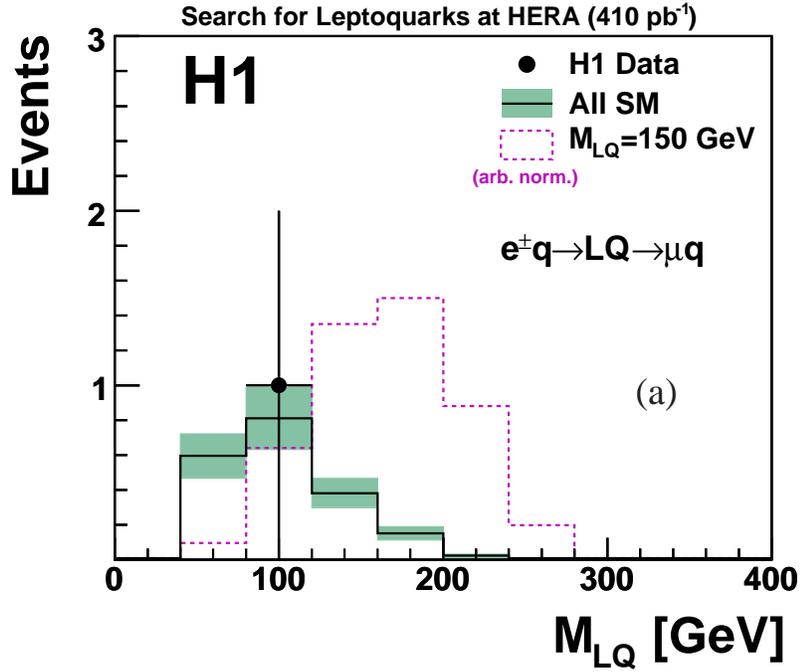}     
   \includegraphics[width=0.7\textwidth]{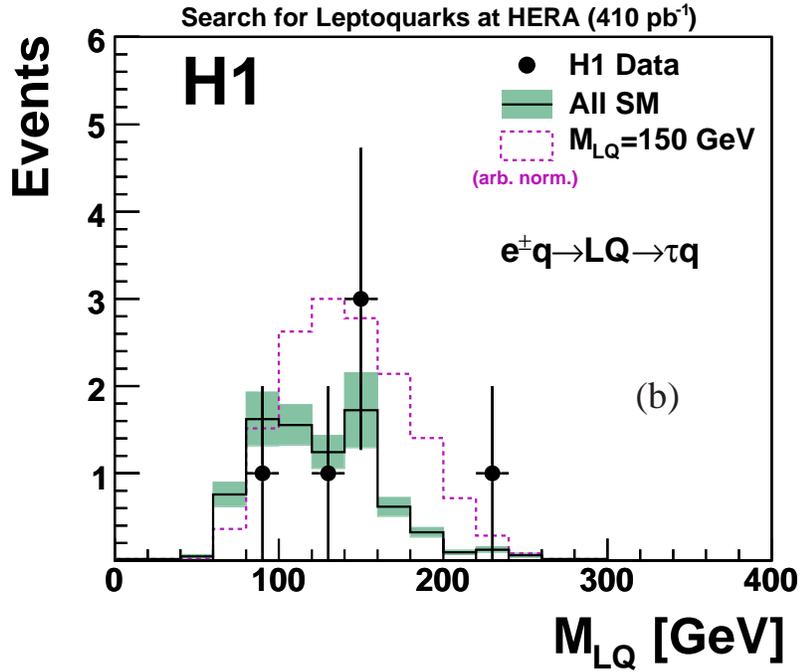}     
  \end{center}   
  \begin{picture} (0.,0.)
    \setlength{\unitlength}{1.0cm}
    \put (11,14){\large{(a)}} 
    \put (11,4.5){\large{(b)}} 
  \end{picture} 
  \vspace{-0.5cm} 
    \caption{The reconstructed leptoquark mass in the search for
    (a)~$ep \rightarrow \mu X$ and (b)~$ep \rightarrow \tau X$
    events. The data are the points and the total uncertainty on the
    SM expectation (open histogram) is given by the shaded band. The
    dashed histogram indicates the LQ signal with arbitrary
    normalisation for a leptoquark mass of $150$~GeV.}
  \label{fig:massplots}
\end{figure}

The observed number of events is in agreement with the SM prediction and therefore no
evidence for LFV is found.
The reconstructed leptoquark mass in the search for $ep \rightarrow \mu X$ and
$ep \rightarrow \tau X$ events is shown in figure~\ref{fig:massplots}, compared to
the SM prediction and an example LQ signal with arbitrary normalisation. 

%%%

In the absence of a signal, the results of the search are interpreted in terms of
exclusion limits on the mass and the coupling of LQs that may mediate LFV. 
The LQ production mechanism at HERA involves non-zero coupling to the
first generation fermions \mbox{$\lambda_{eq} > 0$}.
For the LFV leptoquark decay, it is assumed that only one of the couplings $\lambda_{\mu q}$
and $\lambda_{\tau q}$ is non-zero and that $\lambda_{eq} = \lambda_{\mu q}(\lambda_{\tau q})$,
which results in $\beta_{LFV} = 0.5$.
A modified frequentist method with a likelihood ratio as the test statistic is used to combine 
the individual data sets and the $ep \rightarrow \tau X$ search channels~\cite{Barate:2003sz}.
The lepton beam polarisation enters the limit calculation for the 2003-2007 data.

%%%

\begin{figure}[] 
  \begin{center}
    \includegraphics[width=0.495\textwidth]{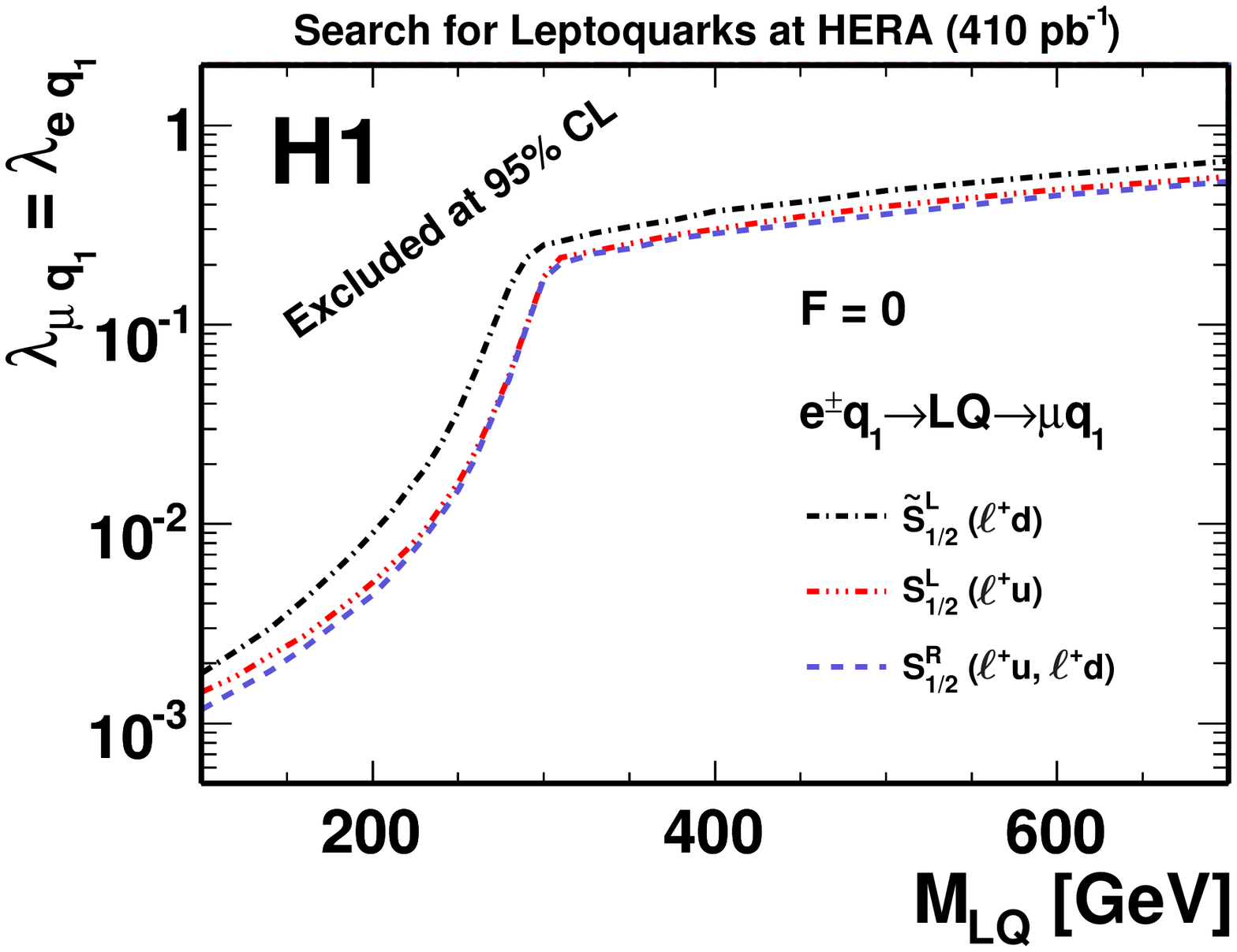}     
    \includegraphics[width=0.495\textwidth]{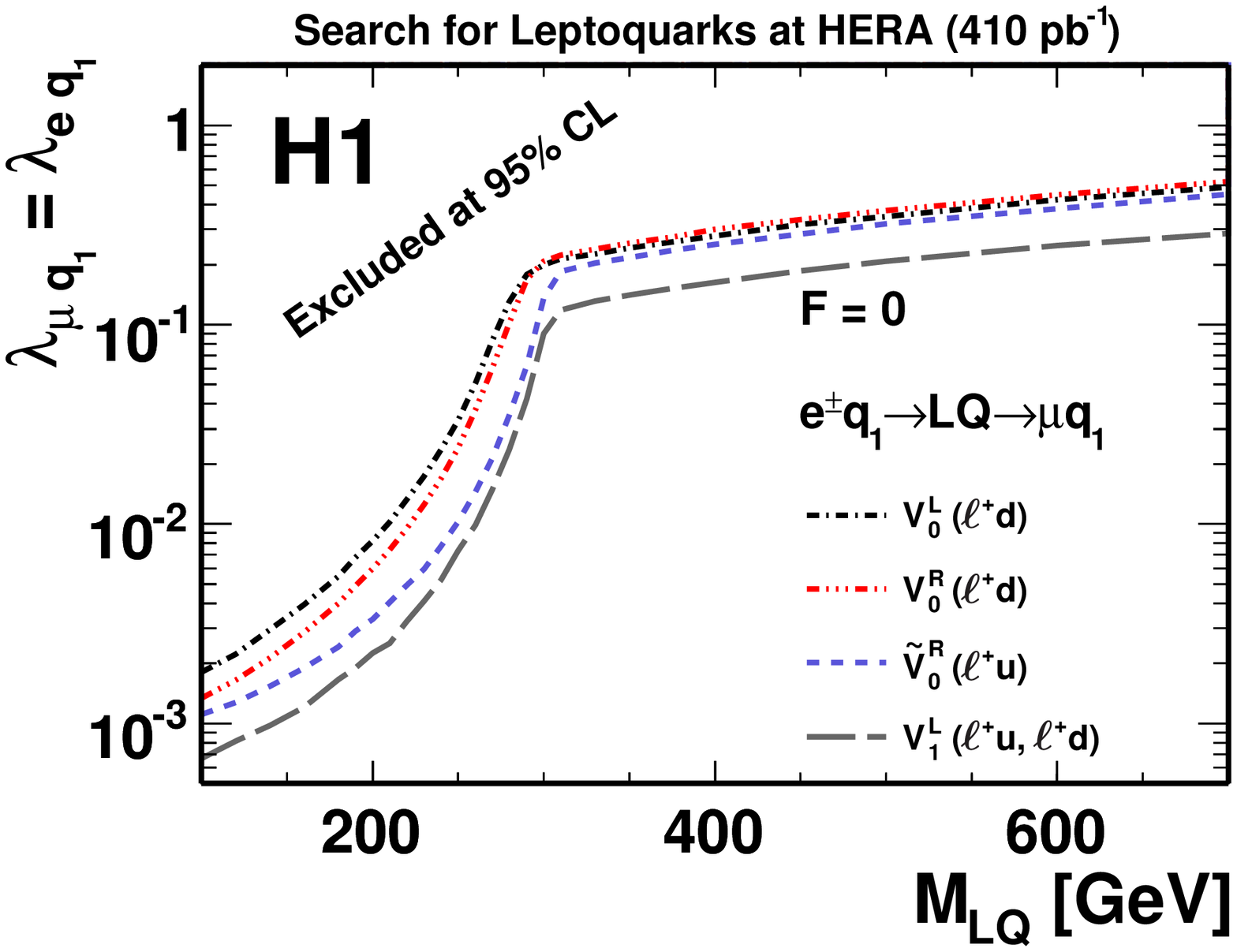}     
    \includegraphics[width=0.495\textwidth]{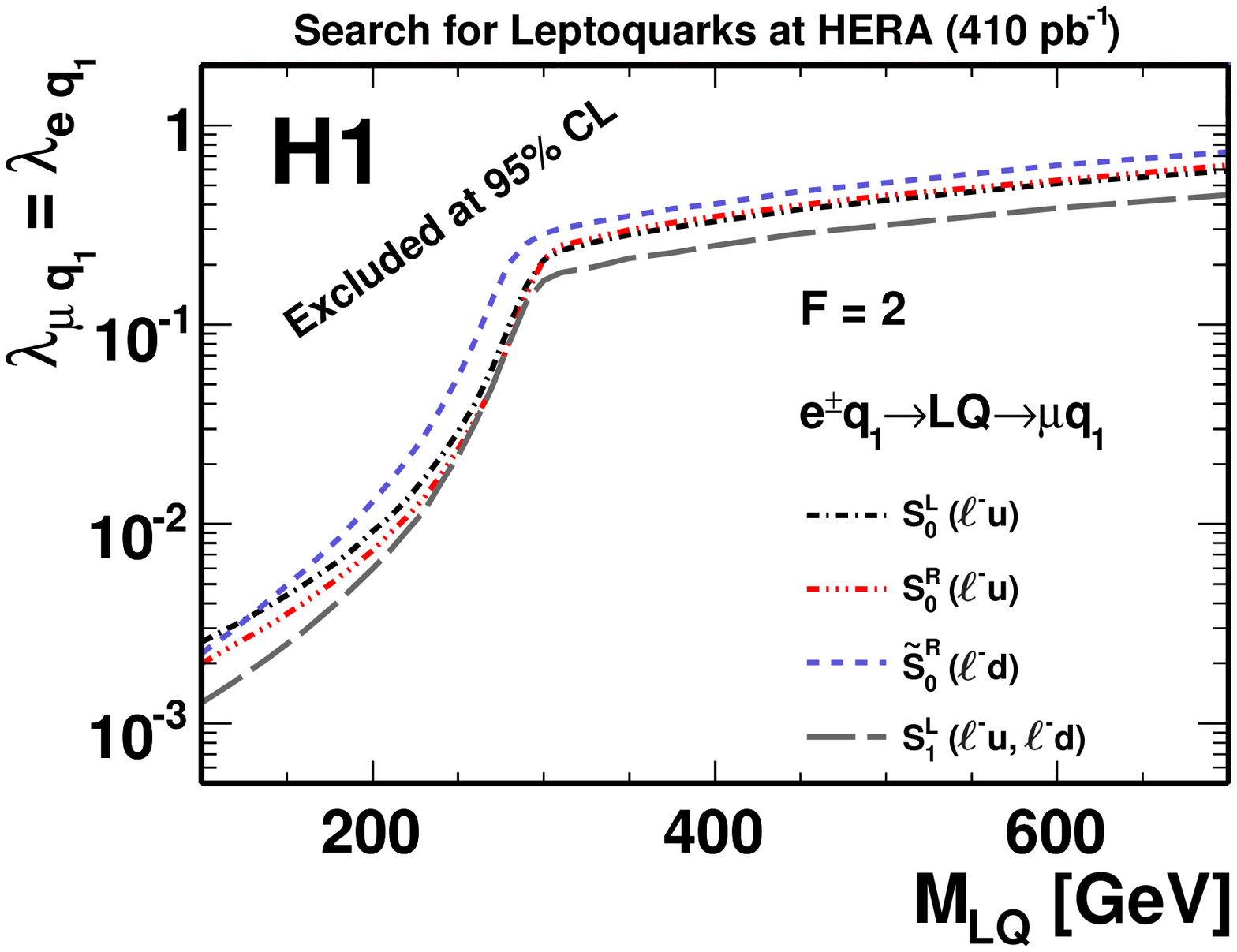}     
    \includegraphics[width=0.495\textwidth]{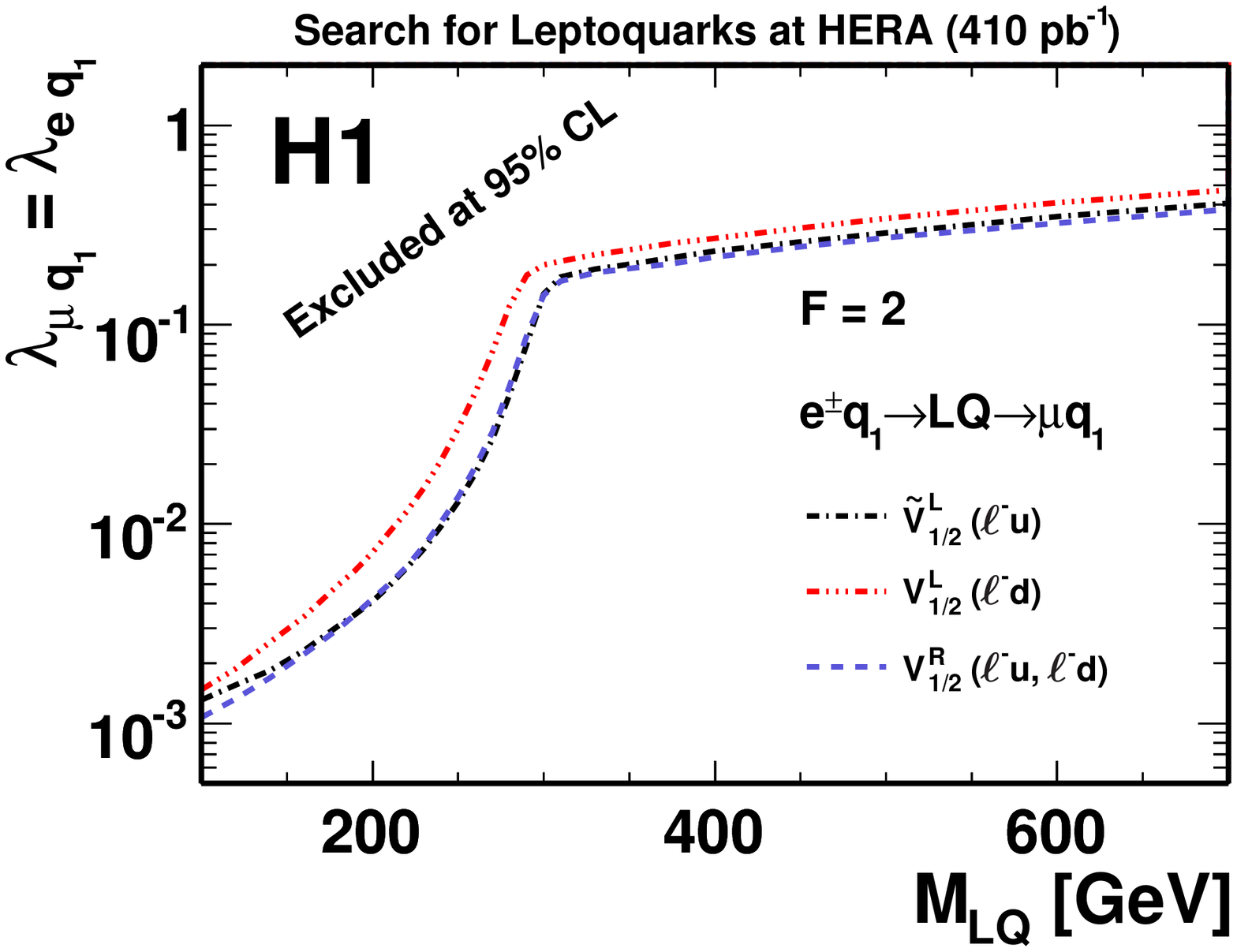}     
  \end{center}   
  \begin{picture} (0.,0.)
    \setlength{\unitlength}{1.0cm}
    \put (3.2,8.7){\large{(a)}} 
    \put (11.2,8.7){\large{(b)}} 
    \put (3.2,3.0){\large{(c)}} 
    \put (11.2,3.0){\large{(d)}} 
  \end{picture} 
  \vspace{-0.5cm} 
    \caption{Exclusion limits on the coupling constants $\lambda_{\mu q} = \lambda_{eq}$ as a function of the leptoquark mass ${\rm M_{LQ}}$
    for (a) scalar LQs with ${\rm F}=0$, (b) vector LQs with ${\rm F}=0$, (c) scalar LQs with ${\rm F}=2$ and (d) vector LQs with ${\rm F}=2$.
    Regions above the lines are excluded at $95\%$~CL.
    The notation $q_1$ indicates that only processes involving first generation quarks are considered.
    The parentheses after the LQ name indicate the fermion pairs coupling to the LQ, where pairs involving anti-quarks are not shown.}
  \label{fig:muonlimits}
\end{figure}

\begin{figure}[] 
  \begin{center}
    \includegraphics[width=0.495\textwidth]{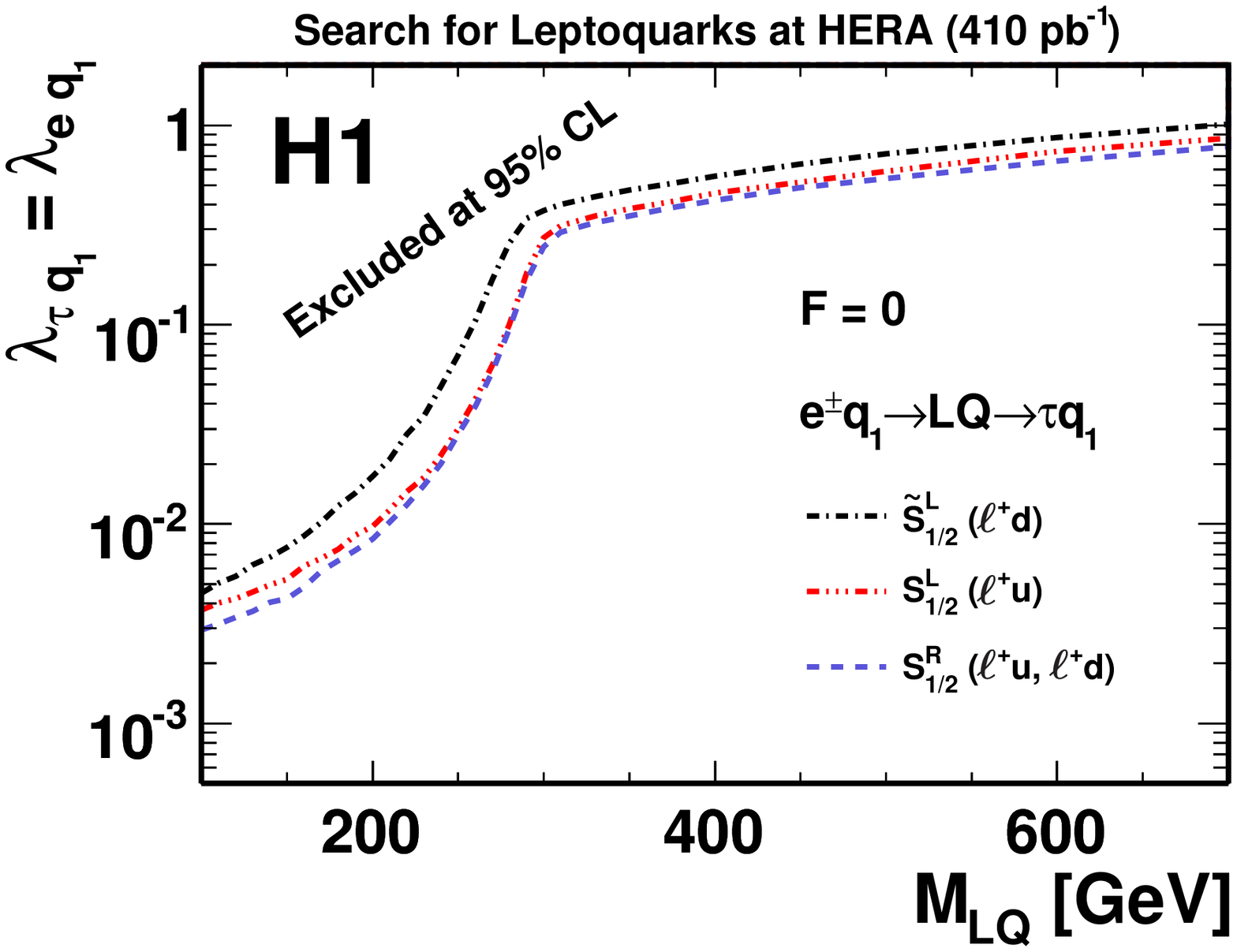}     
    \includegraphics[width=0.495\textwidth]{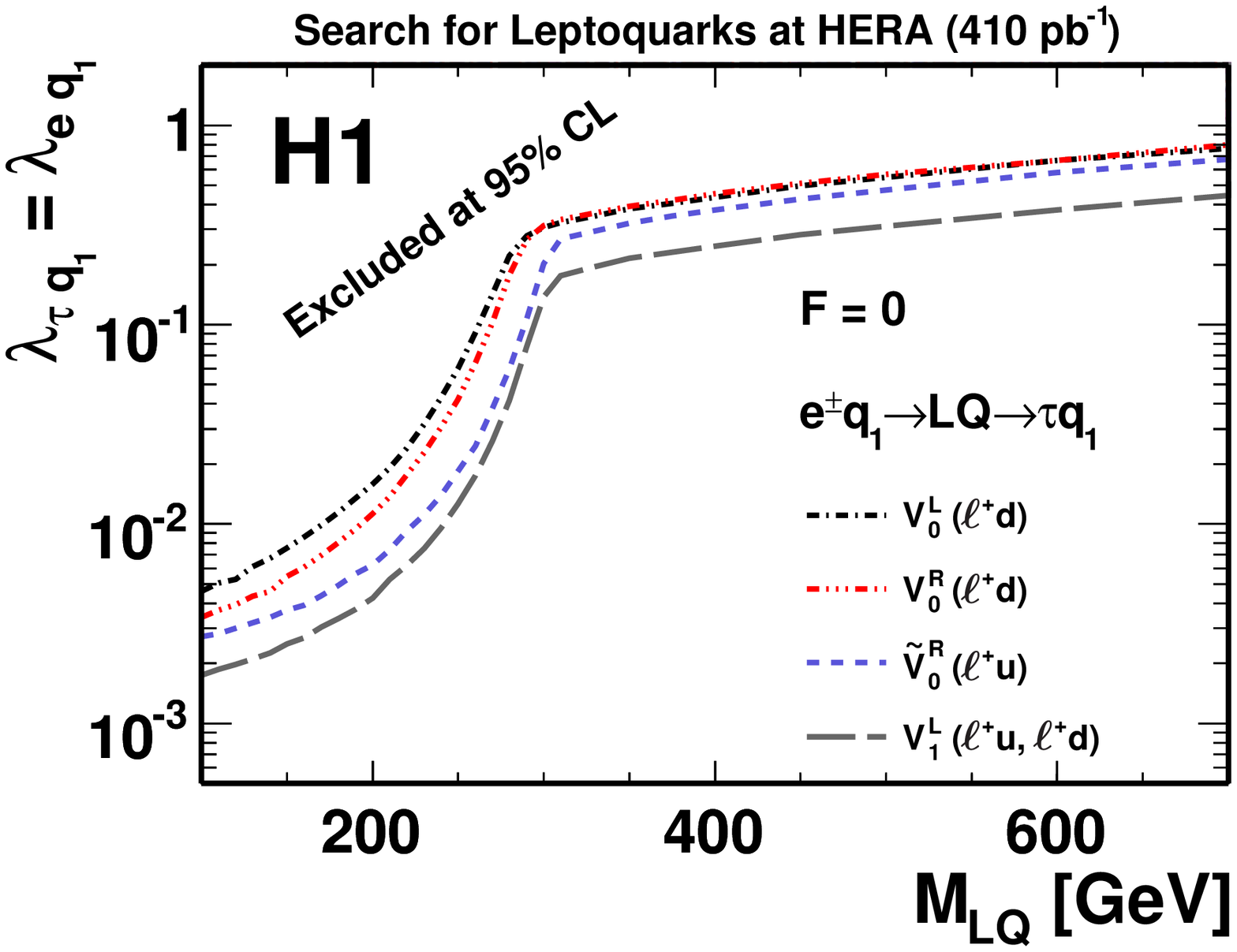}     
    \includegraphics[width=0.495\textwidth]{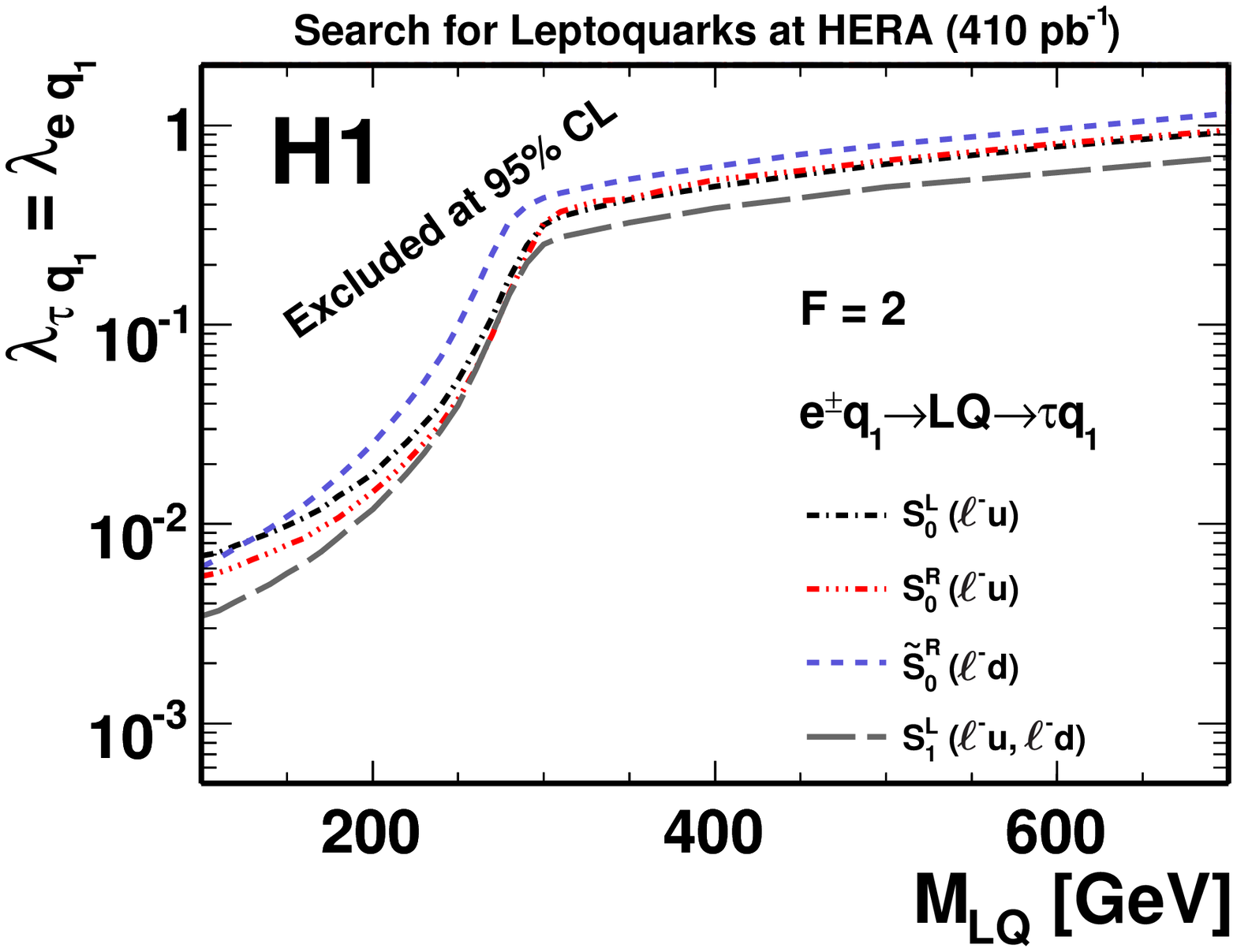}     
    \includegraphics[width=0.495\textwidth]{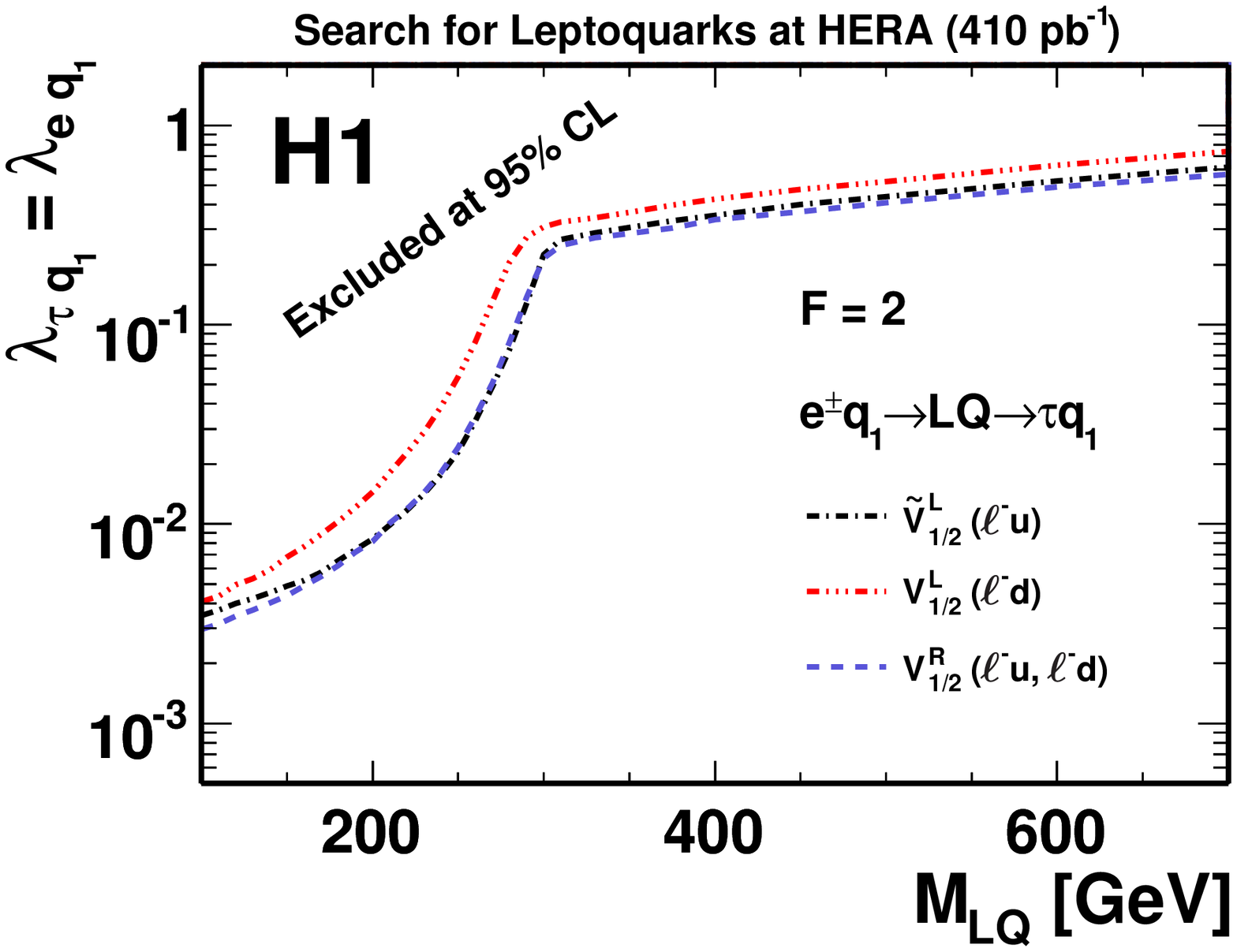}     
 \end{center}   
  \begin{picture} (0.,0.)
    \setlength{\unitlength}{1.0cm}
    \put (3.2,8.7){\large{(a)}} 
    \put (11.2,8.7){\large{(b)}} 
    \put (3.2,3.0){\large{(c)}} 
    \put (11.2,3.0){\large{(d)}} 
  \end{picture} 
  \vspace{-0.5cm} 
    \caption{Exclusion limits on the coupling constants $\lambda_{\tau q} = \lambda_{eq}$ as a function of the leptoquark mass ${\rm M_{LQ}}$
    for (a) scalar LQs with ${\rm F}=0$, (b) vector LQs with ${\rm F}=0$, (c) scalar LQs with ${\rm F}=2$ and (d) vector LQs with ${\rm F}=2$.
    Regions above the lines are excluded at $95\%$~CL.
    The notation $q_1$ indicates that only processes involving first generation quarks are considered.
    The parentheses after the LQ name indicate the fermion pairs coupling to the LQ; pairs involving anti-quarks are not shown.}
  \label{fig:taulimits}
\end{figure}

Figures \ref{fig:muonlimits} and \ref{fig:taulimits} display the $95\%$ confidence level (CL)
upper limits on the coupling $\lambda_{\mu q}$ and $\lambda_{\tau q}$ of all $14$ LQ types
to a muon-quark pair and a tau-quark pair, respectively, as a function of the mass of the LQ
leading to LFV in $ep$ collisions.
Only first generation quarks are considered in these limits.
The limits are most stringent at low LQ masses with values of $\mathcal{O}(10^{-3})$
at $m_{\rm LQ}=100$~GeV.
The limits corresponding to LQs coupling to a $u$ quark are more stringent than
those corresponding  to LQs coupling to the $d$ quark only, as expected from the
larger $u$ quark density in the proton.
Corresponding to the steeply falling parton density function for high values of $x$,
the LQ production cross section decreases rapidly and exclusion limits are less stringent
towards higher LQ masses.
For LQ mass values near the kinematic limit of $319$~GeV, the limit corresponding
to a resonantly  produced LQ turns smoothly into a limit on the virtual effects of both an
off-shell $s$-channel LQ process and a $u$-channel LQ exchange.
At masses $m_{\rm LQ} \gg \sqrt{s}$ the two processes contract to an effective
four-fermion interaction, where the cross section is proportional to 
$(\lambda_{eq}\lambda_{\mu(\tau)q}/m^2_{\rm LQ})^2$.
For a coupling $\lambda$ of electromagnetic strength, where
$\lambda = \sqrt{4 \pi \alpha_{\rm em}} = 0.3$, LFV leptoquarks produced in $ep$
collisions decaying to a muon-quark or a tau-quark pair are excluded at
$95\%$ confidence level up to leptoquark masses of $712$~GeV and
$479$~GeV, respectively.

%%%

The limits on \mbox{$\lambda_{eq}\!=\!\lambda_{\mu(\tau)q}$} in the region
$m_{\rm LQ} \gg \sqrt{s}$ are transformed into a limit on the value 
$\lambda_{eq_i}\lambda_{\mu(\tau)q_j}/m^2_{\rm LQ}$ and shown in tables 
\ref{F0muHighMassLQ} and \ref{F0tauHighMassLQ} for $F=0$ LQs and in 
tables \ref{F2muHighMassLQ} and \ref{F2tauHighMassLQ}  for $F=2$ LQs.
For each LQ type, the limit is calculated for the hypothesis of a process 
with only the quarks of flavours $i$ and $j$ involved.
With respect to quark flavours, the selection criteria described in sections
\ref{sec:secondgen} and \ref{sec:thirdgen} are inclusive since no flavour
tagging of the hadronic jet is used. 
Nevertheless, the sensitivity of the analysis to different quark flavours varies
due to the parton content of the proton and the presence of the $u$-channel
exchange.
Leptoquark couplings to the top quark are not considered in these limits.

%%%

The H1 limits may be compared with constraints from low energy experiments,
based on the non-observation of LFV in muon scattering and rare decays of mesons and 
leptons~\cite{PDG}.
The interpretation of these results in terms of leptoquark exchange and limits on
$\lambda_{eq}\lambda_{\mu(\tau)q_j}/m^2_{\rm LQ}$ \cite{Davidson:1993qk} are also shown
in the tables.
Superior limits are observed by H1 in the search for third generation leptoquarks, compared
to limits from $B \rightarrow \tau \bar{e}$ decays, as well as in a few unique channels.

%%%

At hadron colliders, LQs are mainly produced in pairs independently of $\lambda$,
and therefore searches cannot constrain the LFV couplings.
Lower mass limits by the CMS experiment on second generation scalar leptoquarks
extend up to $394$~GeV~\cite{Khachatryan:2010mq} for a branching ratio $\beta =1$.
Third generation scalar (vector) leptoquarks are ruled out below $247$~GeV by the
D{\O} experiment~\cite{Abazov:2010wq} ($317$~GeV by the
CDF experiment~\cite{Aaltonen:2007rb}) for $\beta =1$.
For $\beta =0.5$, which is a more appropriate value for a comparison to the production
of LQs at HERA, the above second generation search rules out leptoquark
masses below around $300$~GeV, at which mass this analysis rules out
such scalar LQs with couplings in the range $\lambda = 0.2$-$0.3$.

%%%%%%%%%%%%%%%%%%%%%%%%%%%%%%%%%%%%%%%%%%%%%%%%%%%%%%%%%%%%
\section{Conclusion}
\label{sec:con}

A search for lepton flavour violation processes induced by leptoquarks in $ep$ collisions
at a centre-of-mass energy of $319\,{\rm GeV}$ with the H1 experiment at HERA is
presented.
No signal for the LFV processes $ep\rightarrow\mu X$ or $ep\rightarrow\tau X$ 
is observed and assuming a coupling strength of $\lambda=0.3$, leptoquarks mediating
lepton flavour violation are ruled out up to masses of $712$~GeV and $479$~GeV, respectively.
The new H1 limits extend beyond the domain in LQ mass excluded by previous searches
at HERA.
Additionally, the H1 limits remain competitive in certain channels with those from low energy
experiments and for large values of the couplings exclude leptoquark masses beyond
the current limits from hadron colliders.

%%%%%%%%%%%%%%%%%%%%%%%%%%%%%%%%%%%%%%%%%%%%%%%%%%%%%%%%%%%%
\section*{Acknowledgements}

We are grateful to the HERA machine group whose outstanding efforts have made this experiment
possible. We thank the engineers and technicians for their work in constructing and maintaining the
H1 detector, our funding agencies for financial support, the DESY technical staff for continual
assistance and the DESY directorate for support and for the hospitality which they extend to the
non-DESY members of the collaboration.

%%%%%%%%%%%%%%%%%%%%%%%%%%%%%%%%%%%%%%%%%%%%%%%%%%%%%%%%%%%%

\clearpage

\begin{table}
\begin{center}
\setlength{\extrarowheight}{3pt}
\begin{tiny}

\begin{tabular}{|c||c|c|c|c|c|c|c|} 

\hline
\multicolumn{2}{|c}{\rule[-1.8mm]{0mm}{7mm}\Large{$ep \rightarrow \mu X$}}
& \multicolumn{4}{c}{\rule[-1.8mm]{0mm}{7mm}\Large{{\bf H1}}}
& \multicolumn{2}{c|}{\rule[-1.8mm]{0mm}{7mm}\Large{$F=0$}} \\

\hdick
\multicolumn{8}{|c|}{\rule[-1.8mm]{0mm}{7mm}\large{Upper exclusion 
  limits on $\lambda_{eq_i}\lambda_{\mu q_j}/m_{\rm LQ}^2~({\rm TeV}^{-2})$}}\\
\multicolumn{8}{|c|}{\rule[-1.8mm]{0mm}{7mm}\large{for lepton flavour
    violating leptoquarks at $95\%$ CL}}\\

\hline
  & \rule[-1.0mm]{0mm}{7mm}\large{$S_{1/2}^{L}$}  &
 \large{$S_{1/2}^{R}$}  & \large{$\tilde{S}_{1/2}^{L}$} &
 \large{$V_{0}^{L}$}  & \large{$V_{0}^{R}$}   & \large{$\tilde{V}_{0}^{R}$} & \large{$V_{1}^{L}$} \\
\raisebox{1.8ex}[-1.8ex]{\small $q_iq_j$ }
  & $\ell^{-}\bar{U}$ & $\ell^{-} \bar{U}, \ell^{-} \bar{D}$ &    $\ell^{-} \bar{D}$     & $\ell^{-} \bar{D}$ & $\ell^{-} \bar{D}$ & $\ell^{-} \bar{U}$ & $\ell^{-} \bar{U}, \ell^{-} \bar{D}$ \\
  & $\ell^{+}U$ & $\ell^{+} U, \ell^{+} D$ &    $\ell^{+} D$     & $\ell^{+} D$ & $\ell^{+} D$ & $\ell^{+} U$ & $\ell^{+} U, \ell^{+} D$ \\

\hline 
\hline

 & $\mu N \rightarrow eN$ & $\mu N \rightarrow eN$ & $\mu N \rightarrow eN$ & $\mu N \rightarrow eN$ & $\mu N \rightarrow eN$ & $\mu N \rightarrow eN$ & $\mu N \rightarrow eN$  \\
  \small{1~1}  & $5.2 \times 10^{-5}$ & $2.6 \times 10^{-5}$ & $5.2 \times 10^{-5}$ & $2.6 \times 10^{-5}$ & $2.6 \times 10^{-5}$ & $2.6 \times 10^{-5}$ & $0.8 \times 10^{-5}$ \\
 & \bf \normalsize{0.6} & \bf \normalsize{0.6} & \bf \normalsize{0.9} & \bf \normalsize{0.5} & \bf \normalsize{0.6} & \bf \normalsize{0.4} & \bf \normalsize{0.2} \\

\hline

 & $D \rightarrow \mu \bar{e}$ & $K \rightarrow \mu \bar{e}$ & $K \rightarrow \mu \bar{e}$ & $K \rightarrow \mu \bar{e}$ & $K \rightarrow \mu \bar{e}$ & $D \rightarrow \mu \bar{e}$ & $K \rightarrow \mu \bar{e}$  \\
 \small{1~2}   & $0.8$ & $2 \times 10^{-5}$ & $2 \times 10^{-5}$ & $1 \times 10^{-5}$ & $1 \times 10^{-5}$ & $0.4$ & $1 \times 10^{-5}$ \\
 & \bf \normalsize{0.7} \cellcolor{orange} & \bf \normalsize{0.5} & \bf \normalsize{0.9} & \bf \normalsize{0.6} & \bf \normalsize{0.7} & \bf \normalsize{0.5} & \bf \normalsize{0.2} \\

\hline

 & & $B \rightarrow \mu \bar{e}$ & $B \rightarrow \mu \bar{e}$ & $B \rightarrow \mu \bar{e}$ & $B \rightarrow \mu \bar{e}$ &  & $B \rightarrow \mu \bar{e}$   \\
  \small{1~3}   & {\bf \large{$\ast$}} & $0.08$ & $0.08$ & $0.04$ & $0.04$ & {\bf \large{$\ast$}} & $0.04$ \\
 & \bf \normalsize{} & \bf \normalsize{1.0} & \bf \normalsize{0.9} & \bf \normalsize{0.7} & \bf \normalsize{0.8} & \bf \normalsize{} & \bf \normalsize{0.7} \\

\hline
\hline

 & $D \rightarrow \mu \bar{e}$ & $K \rightarrow \mu \bar{e}$ & $K \rightarrow \mu \bar{e}$ & $K \rightarrow \mu \bar{e}$ & $K \rightarrow \mu \bar{e}$ & $D \rightarrow \mu \bar{e}$ & $K \rightarrow \mu \bar{e}$  \\
 \small{2~1}  & $0.8$ & $2 \times 10^{-5}$ & $2 \times 10^{-5}$ & $1 \times 10^{-5}$ & $1 \times 10^{-5}$ & $0.4$ & $1 \times 10^{-5}$ \\
 & \bf \normalsize{1.4} & \bf \normalsize{1.2} & \bf \normalsize{1.5} & \bf \normalsize{0.6} & \bf \normalsize{0.7} & \bf \normalsize{0.5} & \bf \normalsize{0.2} \\

\hline

 & $\mu N \rightarrow eN$ & $\mu N \rightarrow eN$ & $\mu N \rightarrow eN$ & $\mu N \rightarrow eN$ & $\mu N \rightarrow eN$ & $\mu N \rightarrow eN$ & $\mu N \rightarrow eN$  \\
  \small{2~2}  & $9.2 \times 10^{-4}$  & $1.3 \times 10^{-3}$ & $3 \times 10^{-3}$ & $1.5 \times 10^{-3}$ & $1.5 \times 10^{-3}$ & $4.6 \times 10^{-4}$ & $2.7 \times 10^{-4}$ \\
 & \bf \normalsize{2.4} & \bf \normalsize{1.7} & \bf \normalsize{1.9} & \bf \normalsize{1.0} & \bf \normalsize{1.1} & \bf \normalsize{1.4} & \bf \normalsize{0.5} \\

\hline

 & & $B \rightarrow \bar{\mu} eK$ & $B \rightarrow \bar{\mu} eK$ & $B \rightarrow \bar{\mu} eK$ & $B \rightarrow \bar{\mu} eK$ &  & $B \rightarrow \bar{\mu} eK$   \\
 \small{2~3}  & {\bf \large{$\ast$}} & $2.0\times 10^{-3}$ & $2.0\times 10^{-3}$ & $1.0\times 10^{-3}$ & $1.0\times 10^{-3}$ & {\bf \large{$\ast$}} & $1.0\times 10^{-3}$ \\
 & \bf \normalsize{} & \bf \normalsize{2.3} & \bf \normalsize{2.1} & \bf \normalsize{1.4} & \bf \normalsize{1.5} & \bf \normalsize{} & \bf \normalsize{1.4} \\

\hline
\hline

 & & $B \rightarrow \mu \bar{e}$ & $B \rightarrow \mu \bar{e}$ & $V_{ub}$ & $B \rightarrow \mu \bar{e}$ &  & $V_{ub}$  \\
 \small{3~1}  & {\bf \large{$\ast$}} & $0.08$ & $0.08$ & $0.14$ & $0.04$ & {\bf  \large{$\ast$}} & $0.14$ \\
 & \bf \normalsize{} & \bf \normalsize{2.1} & \bf \normalsize{1.9} & \bf \normalsize{0.6} & \bf \normalsize{0.7} & \bf \normalsize{} & \bf \normalsize{0.6} \\

\hline

 &  & $B \rightarrow \bar{\mu}eK$ & $B \rightarrow \bar{\mu}eK$ & $B \rightarrow \bar{\mu}eK$ & $B \rightarrow \bar{\mu}eK$ & & $B \rightarrow \bar{\mu}eK$  \\
 \small{3~2}   & {\bf \large{$\ast$}} & $2.0\times 10^{-3}$ & $2.0\times 10^{-3}$ & $1.0\times 10^{-3}$ & $1.0\times 10^{-3}$ & {\bf \large{$\ast$}} & $1.0\times 10^{-3}$ \\
 & \bf \normalsize{} & \bf \normalsize{3.2} & \bf \normalsize{2.8} & \bf \normalsize{1.1} & \bf \normalsize{1.2} & \bf \normalsize{} & \bf \normalsize{1.1} \\

\hline

 & & $\mu N \rightarrow eN$ & $\mu N \rightarrow eN$ & $\mu N \rightarrow eN$ & $\mu N \rightarrow eN$ &  & $\mu N \rightarrow eN$   \\
 \small{3~3}  & {\bf \large{$\ast$}} & $1.3 \times 10^{-3}$ & $3 \times 10^{-3}$ & $1.5 \times 10^{-3}$ & $1.5 \times 10^{-3}$ & {\bf \large{$\ast$}} & $2.7 \times 10^{-4}$ \\
 & \bf \normalsize{} & \bf \normalsize{3.8} & \bf \normalsize{3.4} & \bf \normalsize{1.7} & \bf \normalsize{1.9} & \bf \normalsize{} & \bf \normalsize{1.7} \\

\hline

\end{tabular}
\end{tiny}
\end{center}
\caption{Limits at 95\% CL on $\lambda_{eq_i}\lambda_{\mu q_j}/m_{\rm LQ}^2$ for $F=0$ leptoquarks (bold).
  The fermion pairs considered in the analysis coupling to each LQ type are indicated in the column headings.
  The $S_{1/2}^{R}$ and $V_{1}^{L}$ LQs couple to both $u$-type ($U$) and $d$-type ($D$) quarks~\cite{BRW}.
  The cases marked with~'$\ast$' refer to scenarios involving a top quark.
  Combinations of $i$ and $j$ shown in the first column denote the quark generation coupling to the electron and muon respectively. 
  In each cell the first two rows show the process providing the most stringent limit  from low energy experiments. 
  Highlighted H1 limits are more stringent than those from the corresponding low energy experiment.}
\label{F0muHighMassLQ}
\end{table}

%%%%%%%%%%%%%%%%%%%%%%%%%%%%%%%%%%%%%%%%%%%%%%%%%%%%%

\begin{table}
\begin{center}
\setlength{\extrarowheight}{3pt}
\begin{tiny}

\begin{tabular}{|c||c|c|c|c|c|c|c|} 

\hline
\multicolumn{2}{|c}{\rule[-1.8mm]{0mm}{7mm}\Large{$ep \rightarrow \tau X$}}
& \multicolumn{4}{c}{\rule[-1.8mm]{0mm}{7mm}\Large{{\bf H1}}}
& \multicolumn{2}{c|}{\rule[-1.8mm]{0mm}{7mm}\Large{$F=0$}} \\

\hdick
\multicolumn{8}{|c|}{\rule[-1.8mm]{0mm}{7mm}\large{Upper exclusion 
  limits on $\lambda_{eq_i}\lambda_{\tau q_j}/m_{\rm LQ}^2~({\rm TeV}^{-2})$}}\\
\multicolumn{8}{|c|}{\rule[-1.8mm]{0mm}{7mm}\large{for lepton flavour
    violating leptoquarks at $95\%$ CL}}\\

\hline
  & \rule[-1.0mm]{0mm}{7mm}\large{$S_{1/2}^{L}$}  &
 \large{$S_{1/2}^{R}$}  & \large{$\tilde{S}_{1/2}^{L}$} &
 \large{$V_{0}^{L}$}  & \large{$V_{0}^{R}$}   & \large{$\tilde{V}_{0}^{R}$} & \large{$V_{1}^{L}$} \\
\raisebox{1.8ex}[-1.8ex]{\small $q_iq_j$ } 
  & $\ell^{-}\bar{U}$ & $\ell^{-} \bar{U}, \ell^{-} \bar{D}$ &    $\ell^{-} \bar{D}$     & $\ell^{-} \bar{D}$ & $\ell^{-} \bar{D}$ & $\ell^{-} \bar{U}$ & $\ell^{-} \bar{U}, \ell^{-} \bar{D}$ \\
  & $\ell^{+}U$ & $\ell^{+} U, \ell^{+} D$ &    $\ell^{+} D$     & $\ell^{+} D$ & $\ell^{+} D$ & $\ell^{+} U$ & $\ell^{+} U, \ell^{+} D$ \\

\hline
\hline

 & $\tau \rightarrow \pi e$ & $\tau \rightarrow \pi e$ & $\tau \rightarrow \pi e$ & $\tau \rightarrow \pi e$ & $\tau \rightarrow  \pi e$ & $\tau \rightarrow \pi e$ & $\tau \rightarrow \pi e$  \\
 \small{1~1} & $0.06$ & $0.03$ & $0.06$ & $0.03$ & $0.03$ & $0.03$ & $0.005$ \\
 & \bf \normalsize{1.4} & \bf \normalsize{1.2} & \bf \normalsize{2.2} & \bf \normalsize{1.2} & \bf \normalsize{1.3} & \bf \normalsize{0.9} & \bf \normalsize{0.4} \\

\hline

 & & $\tau \rightarrow Ke$ & $K \rightarrow \pi \nu \bar{\nu}$ & $\tau \rightarrow Ke$ & $\tau \rightarrow Ke$ &  & $K \rightarrow \pi \nu  \bar{\nu}$  \\
 \small{1~2} & & $0.04$ & $5.8 \times 10^{-4}$ & $0.02$ & $0.02$ &  & $1.5 \times 10^{-4}$ \\
 & \bf \normalsize{1.5} \cellcolor{orange}& \bf \normalsize{1.2} & \bf \normalsize{2.2} & \bf \normalsize{1.5} & \bf \normalsize{1.6} & \bf \normalsize{1.2} & \bf \normalsize{0.5} \\

\hline

 & & $B \rightarrow \tau \bar{e}$ & $B \rightarrow \tau \bar{e}$ & $B \rightarrow \tau \bar{e}$ & $B \rightarrow \tau \bar{e}$ &  & $B \rightarrow \tau \bar{e}$   \\
 \small{1~3} & {\bf \large{$\ast$}} & $0.07$ & $0.07$ & $0.03$ & $0.03$ & {\bf \large{$\ast$}} & $0.03$ \\
 & \bf \normalsize{} & \bf \normalsize{2.2} & \bf \normalsize{2.2} & \bf \normalsize{1.8} & \bf \normalsize{1.8} & \bf \normalsize{} & \bf \normalsize{1.8} \\

\hline
\hline

 & & $\tau \rightarrow Ke$ & $K \rightarrow \pi \nu \bar{\nu}$ & $\tau \rightarrow Ke$ & $\tau \rightarrow Ke$ &  & $K \rightarrow \pi \nu \bar{\nu}$  \\
 \small{2~1} & & $0.04$ & $5.8 \times 10^{-4}$ & $0.02$ & $0.02$ &  & $1.5 \times 10^{-4}$ \\
 & \bf \normalsize{3.4} \cellcolor{orange} & \bf \normalsize{2.8} & \bf \normalsize{3.9} & \bf \normalsize{1.5} & \bf \normalsize{1.6} & \bf \normalsize{1.2} & \bf \normalsize{0.5} \\

\hline

 & $\tau \rightarrow 3 e$ & $\tau \rightarrow 3 e$ & $\tau \rightarrow 3 e$ & $\tau \rightarrow 3 e$ & $\tau \rightarrow 3 e$ & $\tau \rightarrow 3 e$ & $\tau \rightarrow 3 e$  \\
 \small{2~2} & $0.6$ & $0.9$ & $1.8$ & $0.9$ & $0.9$ & $0.3$ & $0.2$ \\
 & \bf \normalsize{6.4} & \bf \normalsize{4.2} & \bf \normalsize{5.0} & \bf \normalsize{2.7} & \bf \normalsize{2.8} & \bf \normalsize{3.5} & \bf \normalsize{1.4} \\

\hline

 & & $B \rightarrow \tau \bar{e}X$ & $B \rightarrow \tau \bar{e}X$ & $B \rightarrow \tau \bar{e}X$ & $B \rightarrow \tau \bar{e}X$ &  & $B \rightarrow \tau \bar{e}X$   \\
 \small{2~3}  & {\bf \large{$\ast$}} & $14.0$ & $14.0$ & $7.2$ & $7.2$ & {\bf \large{$\ast$}} & $7.2$ \\ 
 & \bf \normalsize{} & \bf \normalsize{5.8} \cellcolor{orange} & \bf \normalsize{5.6} \cellcolor{orange} & \bf \normalsize{3.6} \cellcolor{orange} & \bf \normalsize{4.0} \cellcolor{orange} & \bf \normalsize{} & \bf \normalsize{3.6} \cellcolor{orange} \\

\hline
\hline

 & & $B \rightarrow \tau \bar{e}$ & $B \rightarrow \tau \bar{e}$ & $V_{ub}$ & $B \rightarrow \tau \bar{e}$ &  & $V_{ub}$  \\
 \small{3~1} & {\bf \large{$\ast$}} & $0.07$ & $0.07$ & $0.14$ & $0.03$ & {\bf \large{$\ast$}} & $0.14$ \\
 & \bf \normalsize{} & \bf \normalsize{5.3} & \bf \normalsize{4.8} & \bf \normalsize{1.5} & \bf \normalsize{1.7} & \bf \normalsize{} & \bf \normalsize{1.5} \\

\hline

 &  & $B \rightarrow \tau \bar{e}X$ & $B \rightarrow \tau \bar{e}X$ & $B \rightarrow \tau \bar{e}X$ & $B \rightarrow \tau \bar{e}X$ & & $B \rightarrow \tau \bar{e}X$  \\
 \small{3~2} & {\bf \large{$\ast$}} & $14.0$ & $14.0$ & $7.2$ & $7.2$ & {\bf \large{$\ast$}} & $7.2$ \\
 & \bf \normalsize{} & \bf \normalsize{7.9} \cellcolor{orange} & \bf \normalsize{7.6} \cellcolor{orange} & \bf \normalsize{2.9} \cellcolor{orange} & \bf \normalsize{3.1} \cellcolor{orange} & \bf \normalsize{} & \bf \normalsize{2.9} \cellcolor{orange} \\

\hline

 & & $\tau \rightarrow 3 e$ & $\tau \rightarrow 3 e$ & $\tau \rightarrow 3 e$ & $\tau \rightarrow 3 e$ &  & $\tau \rightarrow 3 e$   \\
 \small{3~3} & {\bf \large{$\ast$}} & $0.9$ & $1.8$ & $0.9$ & $0.9$ & {\bf \large{$\ast$}} & $0.2$ \\
 & \bf \normalsize{} & \bf \normalsize{10.1} & \bf \normalsize{9.1} & \bf \normalsize{4.7} & \bf \normalsize{4.9} & \bf \normalsize{} & \bf \normalsize{4.7} \\

\hline

\end{tabular}
\end{tiny}
\end{center}
\caption{Limits at 95\% CL on $\lambda_{eq_i}\lambda_{\tau q_j}/m_{\rm LQ}^2$ for $F=0$ leptoquarks (bold).
  The fermion pairs considered in the analysis coupling to each LQ type are indicated in the column headings.
  The $S_{1/2}^{R}$ and $V_{1}^{L}$ LQs couple to both $u$-type ($U$) and $d$-type ($D$) quarks~\cite{BRW}.
  The cases marked with~'$\ast$' refer to scenarios involving a top quark.
  Combinations of $i$ and $j$ shown in the first column denote the quark generation coupling to the electron and tau lepton respectively. 
  In each cell the first two rows show the process providing the most stringent limit from low energy experiments. 
  Highlighted H1 limits are more stringent than those from the corresponding low energy experiment.}
\label{F0tauHighMassLQ}
\end{table}

%%%%%%%%%%%%%%%%%%%%%%%%%%%%%%%%%%%%%%%%%%%%%%%%%%%%%

\begin{table}
\begin{center}
\setlength{\extrarowheight}{3pt}
\begin{tiny}

\begin{tabular}{|c||c|c|c|c|c|c|c|} 

\hline
\multicolumn{2}{|c}{\rule[-1.8mm]{0mm}{7mm}\Large{$ep \rightarrow \mu X$}}
& \multicolumn{4}{c}{\rule[-1.8mm]{0mm}{7mm}\Large{{\bf H1}}}
& \multicolumn{2}{c|}{\rule[-1.8mm]{0mm}{7mm}\Large{$F=2$}} \\

\hdick
\multicolumn{8}{|c|}{\rule[-1.8mm]{0mm}{7mm}\large{Upper exclusion 
  limits on $\lambda_{eq_i}\lambda_{\mu q_j}/m_{\rm LQ}^2~({\rm TeV}^{-2})$}}\\
\multicolumn{8}{|c|}{\rule[-1.8mm]{0mm}{7mm}\large{for lepton flavour
    violating leptoquarks at $95\%$ CL}}\\

\hline
  & \rule[-1.0mm]{0mm}{7mm}\large{$S_{0}^{L}$}  &
 \large{$S_{0}^{R}$}  & \large{$\tilde{S}_{0}^{R}$} &
 \large{$S_{1}^{L}$}  & \large{$V_{1/2}^{L}$}   & \large{$V_{1/2}^{R}$} & \large{$\tilde{V}_{1/2}^{L}$} \\
\raisebox{1.8ex}[-1.8ex]{\small $q_iq_j$ } 
  & $\ell^{-} U$ & $\ell^{-} U$ &    $\ell^{-} D$  & $\ell^{-} U, \ell^{-} D$   & $\ell^{-} D$ & $\ell^{-} U, \ell^{-} D$ & $\ell^{-} U$ \\
  & $\ell^{+} \bar{U}$ & $\ell^{+} \bar{U}$ &    $\ell^{+} \bar{D}$  & $\ell^{+} \bar{U}, \ell^{+} \bar{D}$   & $\ell^{+} \bar{D}$ & $\ell^{+} \bar{U}, \ell^{+} \bar{D}$ & $\ell^{+} \bar{U}$ \\

\hline
\hline

 & $\mu N \rightarrow eN$ & $\mu N \rightarrow eN$ & $\mu N \rightarrow eN$ & $\mu N \rightarrow eN$ & $\mu N \rightarrow eN$ & $\mu N \rightarrow eN$ & $\mu N \rightarrow eN$  \\
  \small{1~1}  & $5.2 \times 10^{-5}$ & $5.2 \times 10^{-5}$ & $5.2 \times 10^{-5}$ & $1.7 \times 10^{-5}$ & $2.6 \times 10^{-5}$ & $1.3 \times 10^{-5}$ & $2.6 \times 10^{-5}$ \\
 & \bf \normalsize{0.7} & \bf \normalsize{0.8} & \bf \normalsize{1.1} & \bf \normalsize{0.4} & \bf \normalsize{0.5} & \bf \normalsize{0.3} & \bf \normalsize{0.3} \\

\hline

 & $K \rightarrow \pi \nu \bar{\nu}$ & $D \rightarrow \mu \bar{e}$ & $K \rightarrow \mu \bar{e}$ & $K \rightarrow \mu \bar{e}$ & $K \rightarrow \mu \bar{e}$ & $K \rightarrow \mu \bar{e}$ & $D \rightarrow \mu \bar{e}$  \\
 \small{1~2} & $1 \times 10^{-3}$ & $0.8$ & $2 \times 10^{-5}$ & $1 \times 10^{-5}$ & $1 \times 10^{-5}$ & $1 \times 10^{-5}$ & $0.4$ \\
 & \bf \normalsize{0.8} & \bf \normalsize{0.9} & \bf \normalsize{1.2} & \bf \normalsize{0.4} & \bf \normalsize{0.8} & \bf \normalsize{0.5} & \bf \normalsize{0.6} \\

\hline

 & & & $B \rightarrow \mu \bar{e}$ & $V_{ub}$ & $B \rightarrow \mu \bar{e}$ & $B \rightarrow \mu \bar{e}$ &  \\
 \small{1~3} & {\bf \large{$\ast$}} & {\bf \large{$\ast$}} & $0.08$ & $0.3$   & $0.04$ & $0.04$ & {\bf \large{$\ast$}} \\% update 2011 LELIMITS
 & \bf \normalsize{} & \bf \normalsize{} & \bf \normalsize{1.3} & \bf \normalsize{0.6} & \bf \normalsize{0.9} & \bf \normalsize{1.0} & \bf \normalsize{} \\

\hline
\hline

 & $K \rightarrow \pi \nu \bar{\nu}$ & $D \rightarrow \mu \bar{e}$ & $K \rightarrow \mu \bar{e}$ & $K \rightarrow \mu \bar{e}$ & $K \rightarrow \mu \bar{e}$ & $K \rightarrow \mu \bar{e}$ & $D \rightarrow \mu \bar{e}$  \\
 \small{2~1}  & $1 \times 10^{-3}$ & $0.8$ & $2 \times 10^{-5}$ & $1 \times 10^{-5}$ & $1 \times 10^{-5}$ & $1 \times 10^{-5}$ & $0.4$ \\
 & \bf \normalsize{1.2} & \bf \normalsize{1.2} & \bf \normalsize{1.5} & \bf \normalsize{0.6} & \bf \normalsize{0.5} & \bf \normalsize{0.3} & \bf \normalsize{0.4} \cellcolor{orange}  \\

\hline

 & $\mu N \rightarrow eN$ & $\mu N \rightarrow eN$ & $\mu N \rightarrow eN$ & $\mu N \rightarrow eN$ & $\mu N \rightarrow eN$ & $\mu N \rightarrow eN$ & $\mu N \rightarrow eN$  \\
  \small{2~2}  & $9.2 \times 10^{-4}$  & $9.2 \times 10^{-3}$ & $3 \times 10^{-3}$ & $2.5 \times 10^{-3}$ & $1.5 \times 10^{-3}$ & $6.7 \times 10^{-4}$ & $4.6 \times 10^{-4}$ \\
 & \bf \normalsize{2.4} & \bf \normalsize{2.7} & \bf \normalsize{2.1} & \bf \normalsize{0.9} & \bf \normalsize{1.0} & \bf \normalsize{0.9} & \bf \normalsize{1.2} \\

\hline

 & & & $B \rightarrow \bar{\mu} eK$ & $B \rightarrow \bar{\mu} eK$ & $B \rightarrow \bar{\mu} eK$ & $B \rightarrow \bar{\mu} eK$ &  \\
 \small{2~3}  & {\bf \large{$\ast$}} & {\bf \large{$\ast$}} & $2.0 \times 10^{-3}$ & $1.0 \times 10^{-3}$ & $1.0 \times 10^{-3}$ & $1.0 \times 10^{-3}$ & {\bf \large{$\ast$}}\\ % update 2011 LELIMITS
& \bf \normalsize{} & \bf \normalsize{} & \bf \normalsize{2.3} & \bf \normalsize{1.0} & \bf \normalsize{1.4} & \bf \normalsize{1.5} & \bf \normalsize{} \\

\hline
\hline

 & & & $B \rightarrow \mu \bar{e}$ & $B \rightarrow \mu \bar{e}$ & $B \rightarrow \mu \bar{e}$ & $B \rightarrow \mu \bar{e}$ &    \\
 \small{3~1} & {\bf \large{$\ast$}} & {\bf \large{$\ast$}} & $0.08$ & $0.08$ & $0.04$ & $0.04$ & {\bf \large{$\ast$}} \\ % update 2011 LELIMITS
 & \bf \normalsize{} & \bf \normalsize{} & \bf \normalsize{1.8} & \bf \normalsize{0.8} & \bf \normalsize{0.5} & \bf \normalsize{0.5} & \bf \normalsize{} \\

\hline

 & & & $B \rightarrow \bar{\mu}eK$ & $B \rightarrow \bar{\mu}eK$ & $B \rightarrow \bar{\mu}eK$ & $B \rightarrow \bar{\mu}eK$ & \\
 \small{3~2} & {\bf \large{$\ast$}} & {\bf \large{$\ast$}} & $2.0 \times 10^{-3}$ & $1.0\times 10^{-3}$ & $1.0 \times 10^{-3}$ & $1.0 \times 10^{-3}$ & {\bf \large{$\ast$}} \\% update 2011 LELIMITS
 & \bf \normalsize{} & \bf \normalsize{} & \bf \normalsize{3.2} & \bf \normalsize{1.4} & \bf \normalsize{1.1} & \bf \normalsize{1.2} & \bf \normalsize{} \\

\hline

 & & & $\mu N \rightarrow eN$ & $\mu N \rightarrow eN$ & $\mu N \rightarrow eN$ & $\mu N \rightarrow eN$ &    \\
 \small{3~3} & {\bf \large{$\ast$}} & {\bf \large{$\ast$}} & $3 \times 10^{-3}$ & $2.5 \times 10^{-3}$ & $1.5 \times 10^{-3}$ & $6.7 \times 10^{-4}$ & {\bf \large{$\ast$}}  \\
 & \bf \normalsize{} & \bf \normalsize{} & \bf \normalsize{3.8} & \bf \normalsize{1.7} & \bf \normalsize{1.7} & \bf \normalsize{1.9} & \bf \normalsize{} \\

\hline

%%%%%%%%%%%

\end{tabular}
\end{tiny}
\end{center}
\caption{Limits at 95\% CL on $\lambda_{eq_i}\lambda_{\mu q_j}/m_{\rm LQ}^2$ for $F=2$ leptoquarks (bold).
  The fermion pairs considered in the analysis coupling to each LQ type are indicated in the column headings.
  The $S_{1}^{L}$ and $V_{1/2}^{R}$ LQs couple to both $u$-type ($U$) and $d$-type ($D$) quarks~\cite{BRW}.
  The cases marked with~'$\ast$' refer to scenarios involving a top quark.
  Combinations of $i$ and $j$ shown in the first column denote the quark generation coupling to the electron and muon respectively. 
  In each cell the first two rows show the process providing the most stringent limit from low energy experiments. 
  Highlighted H1 limits are comparable to those from the corresponding low energy experiment.}
\label{F2muHighMassLQ}
\end{table}

%%%%%%%%%%%%%%%%%%%%%%%%%%%%%%%%%%%%%%%%%%%%%%%%%%%%%

\begin{table}
\begin{center}
\setlength{\extrarowheight}{3pt}
\begin{tiny}

\begin{tabular}{|c||c|c|c|c|c|c|c|} 

\hline
\multicolumn{2}{|c}{\rule[-1.8mm]{0mm}{7mm}\Large{$ep \rightarrow \tau X$}}
& \multicolumn{4}{c}{\rule[-1.8mm]{0mm}{7mm}\Large{{\bf H1}}}
& \multicolumn{2}{c|}{\rule[-1.8mm]{0mm}{7mm}\Large{$F=2$}} \\

\hdick
\multicolumn{8}{|c|}{\rule[-1.8mm]{0mm}{7mm}\large{Upper exclusion 
  limits on $\lambda_{eq_i}\lambda_{\tau q_j}/m_{\rm LQ}^2~({\rm TeV}^{-2})$}}\\
\multicolumn{8}{|c|}{\rule[-1.8mm]{0mm}{7mm}\large{for lepton flavour
    violating leptoquarks at $95\%$ CL}}\\
\hline
  & \rule[-1.0mm]{0mm}{7mm}\large{$S_{0}^{L}$}  &
 \large{$S_{0}^{R}$}  & \large{$\tilde{S}_{0}^{R}$} &
 \large{$S_{1}^{L}$}  & \large{$V_{1/2}^{L}$}   & \large{$V_{1/2}^{R}$} & \large{$\tilde{V}_{1/2}^{L}$} \\
\raisebox{1.8ex}[-1.8ex]{\small $q_iq_j$ }
  & $\ell^{-} U$ & $\ell^{-} U$ &    $\ell^{-} D$  & $\ell^{-} U, \ell^{-} D$   & $\ell^{-} D$ & $\ell^{-} U, \ell^{-} D$ & $\ell^{-} U$ \\
  & $\ell^{+} \bar{U}$ & $\ell^{+} \bar{U}$ &    $\ell^{+} \bar{D}$  & $\ell^{+} \bar{U}, \ell^{+} \bar{D}$   & $\ell^{+} \bar{D}$ & $\ell^{+} \bar{U}, \ell^{+} \bar{D}$ & $\ell^{+} \bar{U}$ \\

\hline
\hline

 & $G_F$ & $\tau \rightarrow \pi e$ & $\tau \rightarrow \pi e$ & $\tau \rightarrow \pi e$ & $\tau \rightarrow \pi e$ & $\tau \rightarrow \pi e$ & $\tau \rightarrow \pi e$  \\
 \small{1~1} & $0.3$ & $0.06$ & $0.06$ & $0.01$ & $0.03$ & $0.01$ & $0.03$ \\
 & \bf \normalsize{1.6} & \bf \normalsize{1.8} & \bf \normalsize{2.6} & \bf \normalsize{1.0} & \bf \normalsize{1.1} & \bf \normalsize{0.7} & \bf \normalsize{0.8} \\

\hline

 & $K \rightarrow \pi \nu \bar{\nu}$ & & $\tau \rightarrow Ke$ & $K \rightarrow \pi \nu \bar{\nu}$ & $K \rightarrow \pi \nu \bar{\nu}$ & $\tau \rightarrow Ke$ &  \\
 \small{1~2} & $5.8 \times 10^{-4}$ & & $0.04$ & $2.9 \times 10^{-4}$ & $2.9 \times 10^{-4}$ & $0.02$ &  \\
 & \bf \normalsize{1.9} & \bf \normalsize{2.1} \cellcolor{orange}  & \bf \normalsize{2.9} & \bf \normalsize{1.1} & \bf \normalsize{1.9} & \bf \normalsize{1.3} & \bf \normalsize{1.5} \cellcolor{orange} \\

\hline

 & & & $B \rightarrow \tau \bar{e}$ & $V_{ub}$ & $B \rightarrow \tau \bar{e}$ & $B \rightarrow \tau \bar{e}$ &  \\
 \small{1~3} & {\bf \large{$\ast$}} & {\bf \large{$\ast$}} & $0.07$ & $0.3$ & $0.03$ & $0.03$ & {\bf \large{$\ast$}}\\% update 2011 LELIMITS
 & \bf \normalsize{} & \bf \normalsize{} & \bf \normalsize{3.0} & \bf \normalsize{1.3} & \bf \normalsize{2.2} & \bf \normalsize{2.4} & \bf \normalsize{} \\

\hline
\hline

 & $K \rightarrow \pi \nu \bar{\nu}$ & & $\tau \rightarrow Ke$ & $K \rightarrow \pi \nu \bar{\nu}$ & $K \rightarrow \pi \nu \bar{\nu}$ & $\tau \rightarrow Ke$ &  \\
 \small{2~1} & $5.8 \times 10^{-4}$ & & $0.04$ & $2.9 \times 10^{-4}$ & $2.9 \times 10^{-4}$ & $0.02$ &  \\% update 2011 LELIMITS
 & \bf \normalsize{2.7} & \bf \normalsize{2.7} \cellcolor{orange} & \bf \normalsize{3.5} & \bf \normalsize{1.4} & \bf \normalsize{1.2} & \bf \normalsize{0.7} & \bf \normalsize{0.9} \cellcolor{orange} \\

\hline

 & $\tau \rightarrow 3e$ & $\tau \rightarrow 3e$ & $\tau \rightarrow 3e$ & $\tau \rightarrow 3e$ & $\tau \rightarrow 3e$ & $\tau \rightarrow 3e$ & $\tau \rightarrow 3e$  \\
 \small{2~2} & $0.6$  & $0.6$ & $1.8$ & $1.5$ & $0.9$ & $0.5$ & $0.3$ \\% update 2011 LELIMITS
 & \bf \normalsize{6.3} & \bf \normalsize{6.8} & \bf \normalsize{5.4} & \bf \normalsize{2.3} & \bf \normalsize{2.7} & \bf \normalsize{2.2} & \bf \normalsize{3.4} \\

\hline

 & & & $B \rightarrow \bar{\tau} eX$ & $B \rightarrow \bar{\tau} eX$ & $B \rightarrow \bar{\tau} eX$ & $B \rightarrow \bar{\tau} eX$ &  \\
 \small{2~3} & {\bf \large{$\ast$}} & {\bf \large{$\ast$}} & $14.0$ & $7.2$ & $7.2$ & $7.2$ & {\bf \large{$\ast$}}\\
 & \bf \normalsize{} & \bf \normalsize{} & \bf \normalsize{5.8} \cellcolor{orange} & \bf \normalsize{2.7} \cellcolor{orange} & \bf \normalsize{3.6} \cellcolor{orange} & \bf \normalsize{4.0} \cellcolor{orange} & \bf \normalsize{} \\

\hline
\hline

 & & & $B \rightarrow \tau \bar{e}$ & $B \rightarrow \tau \bar{e}$ & $B \rightarrow \tau \bar{e}$ & $B \rightarrow \tau \bar{e}$ &    \\
 \small{3~1}& {\bf \large{$\ast$}} & {\bf \large{$\ast$}} & $0.07$ & $0.03$ & $0.03$ & $0.03$ &  {\bf \large{$\ast$}}\\% update 2011 LELIMITS
 & \bf \normalsize{} & \bf \normalsize{} & \bf \normalsize{4.0} & \bf \normalsize{2.0} & \bf \normalsize{1.2} & \bf \normalsize{1.3} & \bf \normalsize{} \\

\hline

 & & & $B \rightarrow \bar{\tau}eX$ & $B \rightarrow \bar{\tau}eX$ & $B \rightarrow \bar{\tau}eX$ & $B \rightarrow \bar{\tau}eX$ & \\
 \small{3~2} & {\bf \large{$\ast$}} & {\bf \large{$\ast$}} & $14.0$ & $7.2$ & $7.2$ & $7.2$ & {\bf \large{$\ast$}} \\
 & \bf \normalsize{} & \bf \normalsize{} & \bf \normalsize{7.9} \cellcolor{orange} & \bf \normalsize{3.7} \cellcolor{orange} & \bf \normalsize{2.9} \cellcolor{orange} & \bf \normalsize{3.1} \cellcolor{orange} & \bf \normalsize{} \\

\hline

 & & & $\tau \rightarrow 3e$ & $\tau \rightarrow 3e$ & $\tau \rightarrow 3e$ & $\tau \rightarrow 3e$ &    \\
 \small{3~3} & {\bf \large{$\ast$}} & {\bf \large{$\ast$}} & $1.8$ & $1.5$ & $0.9$ & $0.5$ & {\bf \large{$\ast$}}  \\% update 2011 LELIMITS
 & \bf \normalsize{} & \bf \normalsize{} & \bf \normalsize{10.1} & \bf \normalsize{4.6} & \bf \normalsize{4.7} & \bf \normalsize{4.9} & \bf \normalsize{} \\

\hline

\end{tabular}
\end{tiny}
\end{center}
\caption{Limits at 95\% CL on $\lambda_{eq_i}\lambda_{\tau  q_j}/m_{\rm LQ}^2$ for $F=2$ leptoquarks (bold).
  The fermion pairs considered in the analysis coupling to each LQ type are indicated in the column headings.
  The $S_{1}^{L}$ and $V_{1/2}^{R}$ LQs couple to both $u$-type ($U$) and $d$-type ($D$) quarks~\cite{BRW}.
  The cases marked with~'$\ast$' refer to scenarios involving a top quark.
  Combinations of $i$ and $j$ shown in the first column denote the quark generation coupling to the electron and tau lepton respectively.
  In each cell the first two rows show the process providing the most stringent limit from low energy experiments. 
  Highlighted H1 limits are more stringent than those from the  corresponding low energy experiment.}
\label{F2tauHighMassLQ}
\end{table}

%%%%%%%%%%%%%%%%%%%%%%%%%%%%%%%%%%%%%%%%%%%%%%%%%%%%%

%%%%%%%%%%%%%%%%%%%%%%%%%%%%%%%%%%%%%%%%%%%%%%%%%%%%%

%%%%%%%%%%%%%%%%%%%%%%%%%%%%%%%%%%%%%%%%%%%%%%%%%%%%%

\end{document}